\documentclass[ASNA,twocolumn]{USG} 
\usepackage{anyfontsize} %

\usepackage{colortbl}  
\usepackage{xcolor}     

\graphicspath{{./images/}}

\articletype{UNDER REVIEW}%

\usepackage{manyfoot}
\usepackage{amsmath}
\usepackage{amsthm}
\usepackage{tikz}
\usetikzlibrary{calc,patterns,shapes.multipart,arrows,decorations.pathreplacing,fit}
\usepackage{pgfplots}
\usepackage{pgfplotstable}
\pgfplotsset{compat=newest}
\usepackage{hyperref}

\usepackage{booktabs}
\usepackage{multirow}

\usepackage{url}

\usepackage{enumitem}

\begin{document}

\makeatletter
\gdef\@history@dates{} 
\gdef\rhlogo{}         
\renewcommand{\@maketitle}{%
  \hsize=\textwidth\parindent=0\@p@t%
  \thispagestyle{titlepage}%
  \let\footnote\thanks%
  \vskip-1.5mm
  \smash{\@journalfont\@journal}%
  \vskip24.36mm%
  \@@articletype%
  \vskip8.6mm%
  \ifx\@title\empty\else{\@title\par}\fi%
  \removelastskip\vskip8\@p@t%
    {\artauthors\par}%
    \removelastskip\vskip8.25\@p@t%
    \removelastskip\vskip19\@p@t%
    \printabstract
    \removelastskip\vskip\adjtitleskip%
}
\def\@@articletype{%
  \rule[-4\@p@t]{4\@p@t}{15\@p@t}%
  \hskip -0\@p@t%
  {\arttypefont\bf\uppercase{\letterspace to 1.15\naturalwidth{\@articletype}}}%
}
\def\oddfoot@titlepage@info{%
  \vbox{%
    \hsize\textwidth%
    \noindent\rule{\textwidth}{.5\@p@t}%
    \vskip2\@p@t%
    {\fontsize{7}{8}\selectfont \hfill {\pagenumfont\thepage\ of \@LastPg}}%
    \vskip10\@p@t%
  }%
}
\makeatother

\title{\textsc{CHRONOS}: A Hardware-Assisted Phase-Decoupled Framework for Secure Federated Learning in IoT}

\author[1]{Hung Dang}

\address{Faculty of Information Technology, Van Lang School of Technology, Van Lang University, Ho Chi Minh City, Vietnam}

\corres{Hung Dang (\email{hung.dk@vlu.edu.vn})}


\abstract[ABSTRACT]{
    Federated learning enables collaborative training on IoT gateways without sharing raw data, yet gradients remain susceptible to inversion attacks.
     Existing Secure Multiparty Computation defenses impose prohibitive communication overhead, exceeding strict IoT latency and energy budgets.
   
    We propose \textsc{CHRONOS}, a hardware-assisted framework that decouples cryptographic setup from active training. During idle windows,
     \textsc{CHRONOS} executes a once-per-epoch server-relayed Diffie-Hellman exchange within an ARM TrustZone enclave, sealing shared secrets and
     distributing Shamir shares to peers. During training, clients mask gradients via a single stream-cipher evaluation and transmit in one round; a
     hardware-backed counter enforces mask freshness. If clients drop mid-round, the server reconstructs masks from peer-held shares ($k \times 32$
     bytes/client), preserving aggregation without round repetition.
   
    Evaluation on a 32-node heterogeneous testbed (Rock Pi 4 and Orange Pi 5) shows that \textsc{CHRONOS} reduces active-phase latency by up to 74\% over synchronous secure aggregation.  It mitigates gradient inversion while maintaining a persistent Secure World footprint under 1.1 KB, independent of
     model dimension and training horizon.
    }

\maketitle

\section{Introduction}
\label{sec:intro}

Gateway-class Internet of Things (IoT) devices are increasingly deployed across various applications such as
smart energy management, healthcare monitoring, and industrial automation. These devices generate 
distributed sensing data whose privacy is of operational consequence. Federated 
learning (FL)~\cite{MCMAHAN} has emerged as a prominent approach for extracting 
intelligence from such data without centralising it. In particular, a population of clients 
collaboratively trains a shared model by exchanging only locally computed gradient 
updates rather than raw measurements. However, privacy preservation remains a 
key concern in such distributed learning environments, for gradient updates 
themselves carry sufficient information to reconstruct private training samples 
under gradient inversion attacks~\cite{IJIS-REVIEW-2026,TIOT-FL-IMMUNE-2025}.  

Nonetheless, privacy-preserving federated learning faces various challenges.
Gradient inversion attacks reconstruct training samples with high fidelity from the
gradient updates a client transmits to the server.
These attacks have grown increasingly potent, escalating from faithful image reconstruction under an honest-but-curious server to the recovery of sensitive attributes against an actively malicious one~\cite{GIDISCUSS,GINVERT,GEMINIO}.
Plaintext gradient exchange is therefore equivalent, from a privacy standpoint, to direct
transmission of training data.

To mitigate these threats, Bonawitz et al. presented Secure Aggregation (SecAgg)~\cite{SECAGG}  construction in which
a client $i$ would apply a pseudorandom mask to their gradients $g_i$ in such a way that a central server can recover
$\sum_i g_i$ from the masked gradients without learning any individual $g_i$. Standard SecAgg establishes the pairwise seeds via an interactive key-agreement sub-protocol
that executes during the active training round, imposing an $O(N^2)$ communication overhead for a population of $N$ clients.
For IoT devices operating under strict per-round latency and energy budgets, this
interactive cost is prohibitive. In fact, a recent survey explicitly identifies synchronous
cryptographic communication during the training phase as the primary barrier to
Privacy-Preserving Machine Learning (PPML) adoption in resource-constrained edge
environments~\cite{SURVEY-ORIGIN,IJIS-SMPC}.

Whereas existing approaches perform this step synchronously, we observe that the pairwise key establishment in SecAgg is \emph{data-independent}:
it requires no knowledge of the gradient and need not happen during the training round. This observation is especially relevant in IoT deployments. In particular, IoT devices alternate between active sensing windows, which have strict latency budgets, and idle windows (maintenance periods, scheduling gaps, or battery charging intervals) where background computation is less constrained.
This temporal structure accommodates a key-agreement phase that runs during the idle window, thereby minimizing communications and computations overhead incurred by the gradient masking in the latency-critical active phase.

We propose \textsc{CHRONOS}, a hardware-assisted framework for private federated learning
that acts on the aforementioned observation.
Specifically, \textsc{CHRONOS} leverages a Trusted Execution Environment (TEE) to ensure OS-level compromise resistance: the system guarantees gradient confidentiality even if the host operating system of the IoT devices is compromised.
In the \emph{idle phase}, each client generates an ephemeral elliptic-curve keypair
\emph{inside} the TrustZone Trusted Application (TA), ensuring that the private key never exists
in Normal World memory. The clients then conduct a  Diffie-Hellman key exchange to establish pairwise masking seeds, which are sealed inside the TA. In addition, each client distributes Shamir secret shares of its ephemeral private key $\mathit{sk}_i$ to its peers to enable dropout recovery.
In the \emph{active phase}, the TA derives the current round's pseudorandom mask from
the stored seeds and a hardware-backed round counter, enabling the client to mask its
gradient and transmit in one communication round in the no-dropout common case. By omitting secondary self-masks, we trade active false-dropout resilience for strict one-round performance, relying on periodic epoch-level key rotation to bound exposure.

This TEE-minimalist choice, wherein the hardware enclave is used only for seed management and freshness enforcement rather than the training process itself, confers three systems-level advantages. It provides hardware-isolated confidentiality, for an adversary with privileged access to the Normal World cannot extract pairwise seeds or unmask future gradients, a guarantee that software-only pre-computation schemes such as our \textsc{CHRONOS-SW} ablation cannot offer. It enforces freshness in hardware, preventing the host OS from rewinding the round counter to reuse masks. Finally, it keeps persistent TEE memory pressure minimal: by permanently shielding only 1016 bytes of key material (for our $N=32$ cohort) rather than storing megabytes of model parameters, \textsc{CHRONOS} eliminates the enclave paging overhead inherent to full-TEE frameworks such as PPFL~\cite{PPFL}. The transient runtime working set scales only with the model dimension and remains well within standard IoT hardware constraints.

In summary, we make the following contributions in this paper:
\begin{itemize}
\item We identify the idle-window structure for IoT device duty cycles and show how it enables a once-per-epoch cryptographic setup phase that eliminates per-round interaction from the latency-critical active phase.

\item We design and implement \textsc{CHRONOS}, a hardware-assisted framework for private federated learning that features constant-size TEE memory footprint regardless of the number of clients participating in the system.

\item We conduct extensive experiments on a 32-node heterogeneous physical testbed comprising Rock Pi 4 and Orange Pi 5 devices to evaluate the performance of our design.
The evaluation demonstrates that \textsc{CHRONOS} achieves hardware-isolated resilience against client-side OS compromise at a modest overhead. Furthermore, it provides an active-phase aggregation latency reduction of up to 74\% for small IoT-representative models compared to traditional, software-only secure aggregation (SecAgg). We provide a rigorous security-gap analysis showing how \textsc{CHRONOS} provides strictly stronger hardware-isolated guarantees than software-only pre-computation solutions (e.g., our \textsc{CHRONOS-SW} ablation) for a modest absolute latency penalty ($\approx$41--99\,ms) to gain genuine rollback resistance, while maintaining a persistent Secure World storage footprint of fewer than 1.1 KB.
\end{itemize}

The rest of this paper is organised as follows.
Section~\ref{sec:background} provides background on federated learning, additive masking technique to enhance privacy in federated learning, the idle-window model, and ARM TrustZone.
Section~\ref{sec:related} reviews prior works.
Section~\ref{sec:design} presents the \textsc{CHRONOS} architecture.
Section~\ref{sec:security} provides the security analysis.
Section~\ref{sec:impl} describes the implementation.
Section~\ref{sec:eval} presents the experimental evaluation.
Section~\ref{sec:discussion} discusses limitations and future work, before Section~\ref{sec:conclusion} concludes our work.

\section{Background}
\label{sec:background}

\subsection{Federated Learning}
\label{sec:bg:fl}

Let us consider a federated learning system comprising a single central
     aggregation server and $N$ distributed clients, where each client operates
     as an Internet of Things (IoT) edge device. Each IoT device $i \in \{1,
     \ldots, N\}$ collects and maintains its own local, privacy-sensitive dataset
     $D_i$, which typically consists of fine-grained sensor measurements. Under
     this architecture, the raw local datasets never leave the physical
     boundaries of their respective devices. Instead, the central server
     orchestrates the collaborative training process by exchanging only model
     parameters and locally computed updates over the network.

The FederatedAveraging Algorithm (FedAvg) proposed by McMahan et al. \cite{MCMAHAN} is one of the most prominent approach for federated learning. In each training round $r$, the central server broadcasts the current global model
     weights $w^{(r)}$ to a  subset of participating clients $S_r$. 
     Each participating client $i$ computes a local model update $g_i$ by performing
     stochastic gradient descent on its own local dataset $D_i$ using $w^{(r)}$
     as the initial weight. The clients then send their computed local updates to
     the server, which aggregate them into a new global model $w^{(r+1)}$:
   
\begin{equation}
      w^{(r+1)} = w^{(r)} - \eta \sum_{i \in S_r} \frac{|D_i|}{\sum_{j \in S_r}
  |D_j|} g_i
  \end{equation} 
  
  in which   $\eta$ is the learning rate.  For simplicity of exposition, the rest of this paper will refer to this
     local update interchangeably as the local gradient $g_i$.

Nonetheless, various works have shown that an adversary observing $g_i$  in plaintext is able 
to derive certain information about its corresponding local dataset $D_i$. 
Gradient inversion attacks could establish pixel-accurate image reconstructions from single-sample gradients via iterative optimisation~\cite{GIDISCUSS}. This threat has since advanced significantly; adaptive attacks exploit the statistical structure of the gradient distribution to tighten reconstructions further under an honest-but-curious threat model~\cite{GINVERT}. Furthermore, if the adversary is upgraded to an actively malicious server capable of crafting malicious global models, language-guided inversion techniques can recover high-fidelity reconstructions and associated sensitive attributes from a single round of gradient exchange~\cite{GEMINIO}.
In IoT settings, where clients transmit gradients trained on fine-grained ambient measurements (occupancy profiles, physiological signals, energy consumption traces), the consequences of a successful inversion are severe.

\subsection{Additive Masking}
\label{sec:bg:smpc}

A standard approach to concealing individual gradients from the aggregation server while
preserving the aggregate is \emph{additive masking with pairwise cancellation}.
Each client $i$ adds a pseudorandom mask $m_i$ to its gradient before transmission, where
the masks are constructed so that they cancel in the sum:
\[
  \sum_{i=1}^{N} m_i = \mathbf{0} \pmod{p}
\]
where $p$ is a quantization modulus chosen to be sufficiently large to
      prevent overflow during aggregation. Since $ \sum_i (g_i + m_i) = \sum_i g_i$, the server 
recovers the correct aggregate gradient without learning any information about individual $g_i$, for each masked value
$(g_i + m_i)$ is pseudorandom from the server's perspective. \\

\textit{Pairwise PRG construction.} 
The standard realisation of this scheme~\cite{SECAGG} assigns a shared pseudorandom
generator (PRG) seed $s_{ij} = s_{ji}$ to each pair of clients $(i, j)$ with $i \neq j$,
established via Diffie-Hellman key agreement.
Client $i$'s mask for round $r$ is:
\begin{equation}
  m_i(r) = \sum_{j > i} \mathrm{PRG}(s_{ij}, r) - \sum_{j < i} \mathrm{PRG}(s_{ji}, r) \pmod{p}
  \label{eq:mask}
\end{equation}
where $\mathrm{PRG}(s, r) \in \mathbb{F}_p^d$ denotes the output of a pseudorandom
generator seeded with $s$ and indexed by round counter $r$, and $d$ is the
 dimension (number of parameters) of the global model.
By construction, $\sum_i m_i(r) = 0$ for all $r$, for each PRG output $\mathrm{PRG}(s_{ij}, r)$
appears with a positive sign in $m_i$ and a negative sign in $m_j$. \\

\textit{Phase-decoupled structure.} 
The computation of $m_i(r)$ requires only the seeds $\{s_{ij}\}_{j \neq i}$ and the
round index $r$; it is data-independent and does not require knowledge 
of the local dataset $D_i$ or the gradient $g_i$.
The key establishment step (the DH key exchange that produces $\{s_{ij}\}$) is
therefore separable from the active training phase.
In particular, this separation is the structural insight that \textsc{CHRONOS} exploits: key establishment
is a once-per-epoch cost that can be executed during any resource-abundant idle period, leaving
the active phase as a single masked-gradient exchange requiring only $O(d)$ work per client.

\subsection{IoT Device Constraints}
\label{sec:bg:iot}

The class of devices targeted by \textsc{CHRONOS} comprises ARM Cortex-A class single-board computers and edge gateways, such as the Rock Pi 4, NVIDIA Jetson Nano, and similar platforms, that aggregate data from local sensor networks in IoT deployments.
These gateway-class devices are characterised by intermittent network connectivity, strict
per-round latency and energy budgets during active sensing, and periods of reduced activity
during which heavier background tasks can be scheduled.

We define the \emph{idle-window model} as a two-state abstraction:

\begin{itemize}
\item \textit{Idle state.} The device is not actively sensing or participating in a
training round.
Computational and network resources are comparatively abundant.
For battery-operated devices, the idle state often coincides with a charging window;
for mains-powered devices (smart meters, industrial gateways), it corresponds to a
scheduled off-peak maintenance window.
The idle state is the natural site for the once-per-epoch DH key establishment.

\item \textit{Active state.} The device is sampling sensors, computing gradients, and
participating in a federated training round.
Latency and energy budgets are strict; background cryptographic tasks must not interfere
with the real-time sensing pipeline.
\end{itemize}

This abstraction is more general than a battery-charging model: it applies to any device
that alternates between high-demand and low-demand operational periods, regardless of power
source.
Specifically, the key insight of \textsc{CHRONOS} is that the DH key
     establishment phase requires $O(N)$ cryptographic operations and network
     exchanges per client. While this is prohibitively expensive during a
     latency-critical active sensing phase, it easily fits within the resource
     budget of a typical idle window. By shifting this burden to the idle state,
     \textsc{CHRONOS} completely eliminates cryptographic interaction from the
     active phase.

\subsection{ARM TrustZone and OP-TEE}
\label{sec:bg:tee}

ARM TrustZone is a hardware-based security architecture that partitions the processor
into two execution environments: the \emph{Normal World} for the host OS and applications,
and the \emph{Secure World} for trusted services~\cite{TZ-PINTO}.
This isolation is enforced at the hardware level, with a dedicated bit in the system bus
(the NS bit) indicating the security state of every memory and peripheral access.
The Secure World maintains its own isolated memory regions and secure peripherals, which are
inaccessible to the Normal World even if the host OS is compromised.

We use OP-TEE (Open Portable Trusted Execution Environment)~\cite{OPTEE}, an open-source Trusted OS
for ARM Cortex-A processors.
The \textsc{CHRONOS} Trusted Application (TA) runs within OP-TEE, leveraging GlobalPlatform
TEE APIs to manage cryptographic keys and perform high-performance mask generation.
It implements the GlobalPlatform TEE Internal Core API, enabling Trusted Applications (TAs)
written in C to execute exclusively in the Secure World.

In \textsc{CHRONOS}, TrustZone serves two roles.
First, it provides \emph{seed secrecy}: the pairwise DH seeds $\{s_{ij}\}$ stored in
Secure World are inaccessible to any Normal World process, including a compromised
operating system.
An adversary who fully owns the Normal World software stack still cannot extract the seeds
and reconstruct the individual masks $m_i(r)$.
Second, it provides \emph{round-counter integrity}: the per-client round counter $C$,
which indexes the PRG and prevents mask reuse, is maintained entirely in the Secure World
and cannot be forged, rewound, or duplicated by a Normal World attacker.

We scope \textsc{CHRONOS}'s implementation and evaluation to Cortex-A
     devices with OP-TEE.
   For Cortex-M microcontrollers, the equivalent framework is Trusted
     Firmware-M (TF-M),
   which provides analogous GlobalPlatform TEE APIs. Because the per-device
     pairwise seed storage
   in \textsc{CHRONOS} scales linearly as $O(N)$ (amounting to only a few
     kilobytes for typical IoT network sizes
   (Section~\ref{sec:design:storage})), the persistent secure-memory footprint
     remains within the strict SRAM budgets of MCU profiles. Porting to TF-M,
   however, is not merely an API adaptation: the architectural transition from
     Armv8-A to Armv8-M replaces the MMU with an MPU, materially altering the
   zero-copy shared-memory marshalling across the secure boundary.
   We discuss this MCU portability constraint further in
     Section~\ref{sec:discussion}.

\section{Related Work}
\label{sec:related}
 We analyze the existing literature by isolating the specific architectural bottlenecks that prior frameworks fail to overcome, thereby positioning \textsc{CHRONOS} within the structural gap of hardware-isolated, phase-decoupled aggregation. \\

\textbf{The Synchronous Cryptographic Bottleneck}
Homomorphic Encryption (HE) and Secure Multiparty Computation (SMPC) represent the prominent approaches to cryptographic PPML. SecureML~\cite{SECUREML} combines HE and garbled circuits, yet HE operations remain prohibitively slow for IoT batch sizes~\cite{IJIS-SMPC}. SecureNN~\cite{SECURENN} improves throughput using arithmetic secret sharing but demands continuous synchronous connectivity, a requirement incompatible with intermittent IoT deployments.

For federated learning specifically, SecAgg~\cite{SECAGG} protects gradients using a pairwise PRG masking construction identical to the core primitive in \textsc{CHRONOS}. However, SecAgg requires synchronous Shamir-share distribution every round, imposing an $O(N^2)$ server communication overhead, presenting a severe bottleneck for latency-critical IoT sensing phases. MicroSecAgg~\cite{MICROSECAGG} reduces round complexity via a single-server model but retains the synchronous key establishment requirement. The SPDZ family (MASCOT~\cite{MASCOT}, MP-SPDZ~\cite{MPSPDZ}) introduced the two-phase cryptographic paradigm via precomputed Beaver triples. However, because FL gradient aggregation is strictly additive, generating multiplication triples is unnecessary and computationally wasteful. \textsc{CHRONOS} replaces OT-based triple generation with a streamlined, once-per-epoch DH key exchange scheduled explicitly during device idle windows. \\

\textbf{The TEE Memory Bottleneck in Edge Environments}
Trusted Execution Environments (TEEs) offer hardware-enforced privacy guarantees. Early frameworks such as PPFL~\cite{PPFL} utilize ARM TrustZone on the client and Intel SGX on the server, employing greedy layer-wise training to mitigate TEE memory limits. However, even with layer-wise training, processing large parameter chunks inside the typically 16 to 32\,MB Secure World of IoT platforms triggers significant paging overheads. GradSec~\cite{GRADSEC} attempts to reduce the trusted computing base by shielding only gradient tensors during the backward pass; nevertheless, it still processes these large structures within the TEE, incurring unacceptable memory pressure.

ARM TrustZone has also been explored for secure data processing and shadow-stack protection in low-end embedded systems~\cite{IJIS-RE-ARM-2020,IJIS-PATTERNS-2025,TIOT-FLASHADOW-2024}, alongside high-performance file sharing and privacy-preserving crowdsensing~\cite{EETSS-SEFS-2025,TIOT-SENSING-PRIVACY-2022}. More broadly, TEEs have been leveraged to secure various distributed architectures, such as establishing fair marketplaces for outsourced computation~\cite{dang2018towards}, enhancing fault tolerance in consensus protocols~\cite{gao2022mixed}, and enabling autonomous membership management for distributed enclaves~\cite{dang2019autonomous}. \textsc{CHRONOS} departs from full-TEE designs by confining enclave use to seed sealing and monotonic counter enforcement. This architectural constraint limits the persistent Secure World footprint to fewer than 1.1 KB, entirely independent of model dimension and training horizon, eliminating memory paging overhead for edge gateways. \\

\textbf{Software-Based Precomputation and Compromise Resistance}
To mitigate synchronous communication costs, One-Shot Secure Aggregation (Hyb-Agg)~\cite{ONESHOTAGG} combines MK-CKKS homomorphic encryption with ECDH-based additive masking. Hyb-Agg achieves one-shot aggregation but does not target client-side TEE isolation or rollback protection; furthermore, it lacks explicit mechanisms to recover aggregated data if clients drop out mid-round without repeating the phase. To evaluate the baseline performance of software-only decoupled architectures, we construct an ablation (\textsc{CHRONOS-SW}) that isolates the phase-decoupled nature of \textsc{CHRONOS} but operates entirely in software, storing seeds and round counters in the Normal World filesystem. We employ \textsc{CHRONOS-SW} as an empirical proxy to quantify the performance parity and security gap of such software-only designs, which provide no protection against a compromised host operating system.
\textsc{CHRONOS} bridges this gap by combining phase decoupling with hardware-isolated confidentiality. By generating the ephemeral private keys inside the TrustZone Trusted Application and sealing the resulting shared secrets, \textsc{CHRONOS} achieves OS-level compromise resistance. An adversary with root access to the Normal World cannot extract the seeds to decrypt future gradients, nor rewind the hardware-backed round counter to reuse masks.

\section{\textsc{CHRONOS} Design}
\label{sec:design}

\subsection{System and Threat Model}
\label{sec:design:threat}

\textbf{System model and notation.} 
Let $N$ be the number of federated learning clients, $D$ the model dimension, $p$ a prime defining the field $\mathbb{F}_p$, $t$ the Shamir reconstruction threshold ($1 \le t \le N-1$), and $\mathcal{P} = \{(i,j) : 1 \le i < j \le N\}$ the set of ordered client pairs. All field arithmetic is performed modulo $p$.

\begin{figure*}[t]
\centering
\resizebox{\textwidth}{!}{\begin{tikzpicture}[
  >=stealth,
  box/.style={draw, rounded corners, align=center, fill=white, thick},
  server/.style={box, fill=blue!10, minimum width=3cm, minimum height=1cm},
  client_small/.style={box, fill=gray!10, minimum size=0.8cm},
  world/.style={box, minimum width=3.5cm, minimum height=3cm},
  normal/.style={world, fill=green!5},
  secure/.style={world, fill=red!5},
  zoom_box/.style={draw, dashed, gray, thick}
]

\node[server] (server) at (0, 6) {\fontsize{8}{9}\selectfont \textbf{Aggregation} \\ \fontsize{8}{9}\selectfont \textbf{Server}};

\node[client_small] (c1) at (-2, 3) {\small $C_1$};
\node[client_small] (c2) at (-0.5, 3) {\small $C_2$};
\node[client_small] (c3) at (1, 3) {\small $C_3$};
\node at (1.8, 3) {\large $\dots$};
\node[client_small] (cn) at (2.5, 3) {\small $C_N$};

\node[font=\bfseries] at (1.5, 2) {IoT Device Cohort ($N$)};

\foreach \c in {c1, c2, c3, cn} {
  \draw[-, gray!40] (\c.north) -- (server.south);
}

\begin{scope}[shift={(10, 3.5)}]
  \node[normal] (nw) at (-2, 0) {\textbf{Normal World}\\\textit{Host OS}\\\\\vspace{2pt}\fontsize{8}{9}\selectfont \textsc{Idle Daemon}\\\fontsize{8}{9}\selectfont \textsc{Training Wrapper}};
  \node[secure] (sw) at (2, 0) {\textbf{Secure World}\\\textit{TrustZone Enclave}\\\\\vspace{2pt}\fontsize{8}{9}\selectfont \textsc{Trusted App (TA)}\\\fontsize{8}{9}\selectfont \textsc{Monotonic Counter} $C$};
  \draw[dashed, thick, gray, rounded corners] ($(nw.north west) + (-0.15, 0.15)$) rectangle ($(sw.south east) + (0.15,-0.15)$);
  
    \draw[ultra thick, dotted, red] (0, 1.8) -- (0, -1.8);
  
  
\end{scope}

\draw[zoom_box] (cn.north east) -- (nw.north west);
\draw[zoom_box] (cn.south east) -- (nw.south west);


\draw[->, thick, blue, rounded corners] (12, 5) |- node[pos=0.4,left, font=\bfseries\scriptsize, text=blue] {(1) Export $pk_i$} ($(server.east)+(0,+0.2)$);

\draw[<-, thick, blue, rounded corners] (7.5, 5) |- node[pos=0.35, font=\bfseries\scriptsize, text=blue, left] {(2) Receive $\{pk_j\}$} ($(server.east)+(0,-0.2)$);

\draw[->, ultra thick, purple, rounded corners] (8.5, 5) -- ++(0, 1.9) -| node[pos=0.1, left, font=\bfseries\scriptsize, text=purple, above] {Transmit $\tilde{g}_i$} (server.north);


\end{tikzpicture}}
\caption{System model of \textsc{CHRONOS}. (Left) Overview of central aggregation server and a $N$-client cohort. (Right) Illustration of hardware architecture of a single client. The diagram maps the temporal phases of the protocol onto the hardware-isolated worlds of ARM TrustZone: the Secure World handles the once-per-epoch cryptographic setup (during Idle Phase, depicted in blue lines), while the Normal World manages the latency-critical gradient transmission (during Active Phase, depicted in purple line) using enclave-derived masks. The vertical red dashed line depicts the hardware isolation between the two worlds.}
\label{fig:architecture}
\end{figure*}
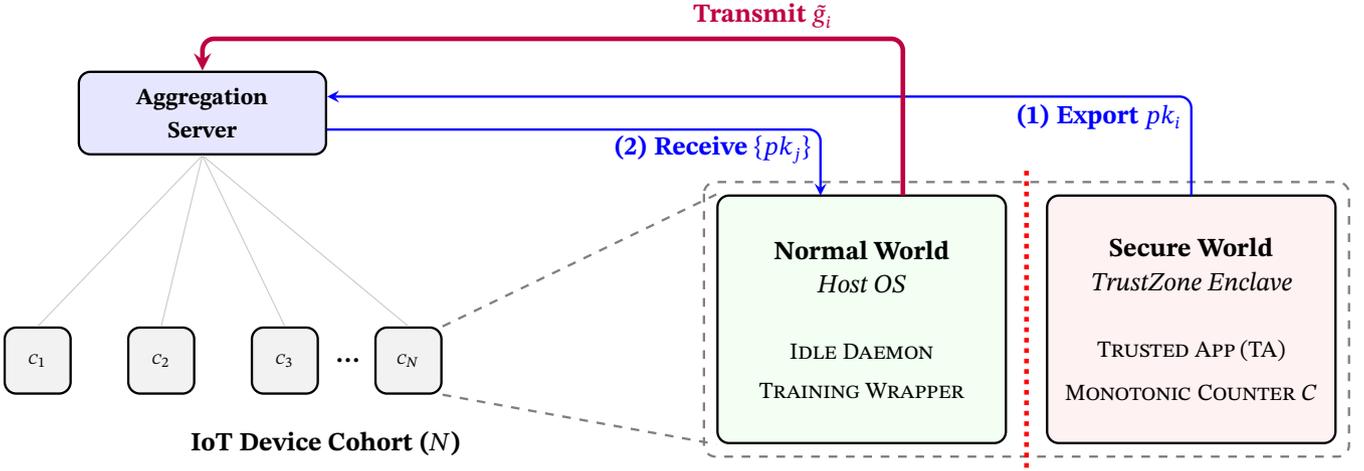

The system, as illustrated in Figure~\ref{fig:architecture}, comprises these $N$ IoT sensing devices (clients) and a central aggregation server.
Each client device features a hardware-partitioned architecture consisting of a Normal World and a Secure World (e.g., ARM TrustZone). The Normal World runs the host operating system, manages the network stack, computes local gradients, and handles general application logic. The Secure World hosts a Trusted Application (TA) with hardware-isolated memory and execution, responsible exclusively for cryptographic setup, key sealing, and mask generation. The server coordinates the training rounds and computes the aggregate of the received masked gradients to update the global model. \\

\textbf{Adversary model.}
We delineate our threat model into two distinct domains: the central aggregation server and the distributed IoT clients.

At the \emph{server level}, the primary adversary is an \emph{honest-but-curious} (HBC) aggregator. The server executes the aggregation and routing protocols faithfully according to the specification. Crucially, the server is trusted not to maliciously manipulate dropout flags (i.e., it will not falsely claim a client has dropped out to trick other clients into revealing secret shares for that client's mask). However, the server attempts to infer private information or reconstruct individual gradients from the masked vectors it receives. We also consider a passive network eavesdropper who observes all client-server communication. Because the server acts as a central relay, it must be prevented from executing Man-in-the-Middle (MITM) attacks during key exchange. We assume the network operator has provisioned each device with a trusted root Certificate Authority (CA) public key during first-time enrollment.

At the \emph{client level}, we assume a much stronger, active adversary. We model a \emph{system-level intrusion} where the adversary has achieved full root-level compromise of the client's Normal World operating system (e.g., via malware or unpatched OS vulnerabilities). In this compromised state, the adversary can intercept all Normal World memory, manipulate network traffic, and attempt to subvert the training process. However, we assume the hardware-backed TrustZone Secure World remains tamper-resistant. The adversary cannot breach the Trusted Application, extract sealed keys from the RPMB, or bypass the hardware-enforced isolation. Physical hardware attacks requiring direct local access to the device (e.g., decapping, microscopic probing) are considered out of scope.

We assume that up to $t-1$ clients may collude with the server or be fully compromised by the adversary. This threshold preserves both the confidentiality of the ephemeral private keys (as $t$ shares are required for Shamir reconstruction) and the secrecy of the training data (as $t \le N-1$ ensures every honest client shares a masking seed with at least one honest peer). \\

\textbf{Security goals.}
\textsc{CHRONOS} targets four primary properties:

\begin{enumerate}
\item \textbf{Data Confidentiality.} The server learns the aggregate gradient
$\sum_i g_i$ but cannot reconstruct any individual client gradient $g_i$.
A passive eavesdropper observing the active-phase traffic learns nothing about any $g_i$.

\item \textbf{OS-Level Compromise Resistance.} The cryptographic guarantees of the system (confidentiality and freshness) must hold even against an adversary who has achieved full root-level compromise of the client's Normal World execution environment.

\item \textbf{Execution Freshness.} Each mask value $m_i(r)$ is consumed exactly once.
The system must prevent a compromised Normal World OS from reusing a mask for two distinct rounds, which would otherwise allow the server to calculate the difference between two gradients ($g_i - g'_i$) and potentially leak private data.

\item \textbf{Functional Correctness and Robustness.} The global model update produced by the server must be
identical to the update that would result from aggregating unmasked gradients. Additionally, the system must degrade gracefully: if a client has not completed key establishment, it drops out of the current round rather than transmitting an unprotected gradient.
\end{enumerate}

\subsection{Overview}
\label{sec:design:overview}

\textsc{CHRONOS} separates the life cycle of a secure federated learning client into two
temporally isolated phases, each mapped to a distinct operational state of the device. \\ 

\textbf{Idle-Phase (key establishment and share distribution).}
Since the device is not engaged in active sensing, computational resources are abundant.
During this phase, the \textsc{CHRONOS} idle daemon activates and orchestrates a
server-relayed Diffie-Hellman key exchange with every other client in the federation.
The ephemeral keypair is generated \emph{inside} the TrustZone Trusted
Application (TA), so the private key $\mathit{sk}_i$ never exists in Normal World memory.
Specifically, the TA computes all pairwise shared secrets, seals them under a device-unique key, and
generates Shamir secret shares of the ephemeral private key $\mathit{sk}_i$ for distribution
to peers; from these shares, any $t$ surviving peers can help the server recover
$\mathit{sk}_i$ and thereby reconstruct the missing client's mask.
The shares (but never the seeds themselves) are exported to the Normal World for relay through the server.
This phase runs exactly once per federation membership change or key-rotation epoch. \\

\textbf{Active-Phase (Mask Generation and Gradient Aggregation).}
The device is operating within a federated training round.
The \textsc{CHRONOS} active-phase wrapper invokes the TA to derive the round-specific mask,
applies it to the local gradient, and transmits the masked gradient in a single
communication round (two rounds under recovery).
If a client drops mid-round, the server requests seed-reconstruction shares from
surviving clients, reconstructs the missing client's mask directly, and corrects the
aggregate without repeating the round.

The following subsections formalise each component.

\subsection{Idle-Phase: Key Establishment, Seed Sealing, and Share Distribution}
\label{sec:design:idle}

\textbf{TEE-internal key generation.}
\textsc{CHRONOS} generates the ephemeral Diffie-Hellman keypair \emph{inside} the TA via a
dedicated GlobalPlatform TEE command (\textsc{KEYGEN}).
The TA invokes the TEE's built-in elliptic-curve key generation routine, exports only the
public key $\mathit{pk}_i$ to the Normal World, and retains the private key $\mathit{sk}_i$ in Secure World
memory.
Because $\mathit{sk}_i$ is generated within and never leaves the hardware boundary, a compromised host OS cannot
extract it, mitigating private-key exposure even if the OS intercepts the TEE API call. \\

\textbf{Authenticated Key Exchange (AKE).}
The central server acts as a common bulletin board to facilitate peer discovery. To mitigate the Man-in-the-Middle (MITM) risk inherent in unauthenticated Diffie-Hellman, where the server could substitute its own public keys, \textsc{CHRONOS} enforces an Authenticated Key Exchange.
During the idle phase, the Normal World daemon sends $\mathit{pk}_i$, which is cryptographically signed by the TA using the device's enrolled certificate, to the server.
The server collects these signed public keys and pushes the peer set $\{\mathit{pk}_j : j \neq i\}$ to each client.
The client daemon passes this set to the TA via a second command
(\textsc{COMPUTE\_SEEDS}). The TA verifies each peer's signature against the trusted root CA. For verified peers, the TA computes the raw Diffie-Hellman secret and applies an HMAC-based Key Derivation Function (HKDF):
\[
  s_{ij} = \mathrm{DH}(\mathit{sk}_i,\, \mathit{pk}_j)
\]
\[
  k^{\text{enc}}_{ij}, k^{\text{prg}}_{ij} = \mathrm{HKDF}(s_{ij})
\]
where $k^{\text{enc}}_{ij}$ is a 128-bit key for share encryption (with freshly generated IV for each encryption) and $k^{\text{prg}}_{ij}$ is a 256-bit key for PRG masking. Distinct info strings are used to derive the two keys from $\mathrm{HKDF}(s_{ij})$.
The symmetry $\mathrm{DH}(\mathit{sk}_i, \mathit{pk}_j) = \mathrm{DH}(\mathit{sk}_j, \mathit{pk}_i)$ ensures these derived keys are identical for both peers.
Once all keys are computed, $\mathit{sk}_i$ is retained in Secure World memory until the
subsequent \textsc{SEAL} call. \\

\textbf{Atomic TEE Sealing and Share Generation.}
The TA completes the idle phase via the \textsc{SEAL}$(t)$ command, which performs four actions with TA-level atomicity to ensure consistency:
\begin{enumerate}
\item The TA generates $t$-of-$(N{-}1)$ secret shares of the ephemeral private key $\mathit{sk}_i$ over byte-wise GF($2^8$), producing $N-1$ shares $\{\sigma_{i \to j}\}_{j \neq i}$. To prevent server interception, it encrypts each share under the corresponding derived encryption key $k^{\text{enc}}_{ij}$ with AES-128-GCM to produce a 60-byte ciphertext $E_{k^{\text{enc}}_{ij}}(\sigma_{i \to j})$ (including a 12-byte IV and 16-byte authentication tag) before writing the ciphertexts to a Normal World buffer for relay to peers.
\item It encrypts the raw Diffie-Hellman shared secret set $\mathcal{S}_i = \{s_{ij}\}_{j \neq i}$ under AES-256-GCM using a key derived from the device Hardware Unique Key (HUK). The resulting ciphertext and authentication tag are written to OP-TEE Secure Storage.
\item It initialises the hardware-backed round counter $C \leftarrow 0$.
\item It securely erases $\mathit{sk}_i$ from Secure World memory.
\end{enumerate}

We note that the atomicity holds for the four TA-local actions; idle-phase completion at the daemon level additionally requires acknowledged delivery of all $N-1$ share ciphertexts to peers.
This ensures that if any step fails, the device remains in a pre-establishment state with no leaked shares or uninitialized counters. Because $\mathit{sk}_i$ is erased immediately and its shares distributed, the device relies exclusively on the sealed seeds for mask generation, while any $t$ peers can assist the server in mask reconstruction if the device drops. \\

\textbf{Share distribution.}
The Normal World daemon then distributes the ciphertext $E_{k^{\text{enc}}_{ij}}(\sigma_{i \to j})$ to peer $j$ for all $j \neq i$,
relayed by the server.
Each client $k$ stores the received ciphertext in ordinary (non-TEE)
local storage for all $j \neq k$.
The $t$-of-$(N{-}1)$ security
guarantee, combined with channel encryption, ensures that any coalition of $t-1$ or fewer clients (along with the server) learns
nothing about $\mathit{sk}_i$ from the shares they hold. \\

\textbf{Storage footprint.}
\label{sec:design:storage}
The TEE stores only the sealed seed set and the round counter:
\[
  \text{TEE storage} = (N-1) \times 32 + 16_{\text{tag}} + 4_{\text{counter}} \;\text{bytes}.
\]
For $N = 32$, the theoretical footprint is 1012\,bytes; the measured value is 1016\,bytes,
reflecting 4\,bytes of OP-TEE session-context overhead not captured in the formula.
Both values are independent of model dimension $D$ and training horizon $R$.
The Normal World additionally stores $N{-}1$ received shares, each 32 bytes (one per
peer): $(N{-}1) \times 32$ bytes of non-sensitive recovery material.

\subsection{Active-Phase: Mask Generation and Gradient Aggregation}
\label{sec:design:active}

\textbf{TA mask generation command.}
The TA exposes two commands during the active phase:
\begin{itemize}
\item \texttt{PEEK\_COUNTER}: returns the current value of $C$ read-only.
\item \texttt{GENERATE\_MASK}$(D, r)$:
atomically reads $\mathcal{S}_i$ from Secure Storage, derives the PRG keys via HKDF,
  validates $r > C$, evaluates Equation~\eqref{eq:mask}, updates $C \leftarrow r$, and returns the mask to the
  Normal World via shared memory.
  If key establishment has not been completed, the command returns
  \texttt{ERR\_NOT\_READY}.
\end{itemize}

\textbf{Active-phase masking.}
When a training round $r$ begins, client $i$ computes its local gradient $g_i \in \mathbb{F}_p^D$.
It invokes \textsc{GenerateMask}$(D, r)$, which evaluates:
\[
  m_i(r) = \sum_{j > i} \mathrm{PRG}(k^{\text{prg}}_{ij},\, r)
          - \sum_{j < i} \mathrm{PRG}(k^{\text{prg}}_{ji},\, r) \pmod{p}.
\]
Here, $\mathrm{PRG}(k^{\text{prg}}, r) \in \mathbb{F}_p^D$ is an AES-128-CTR stream cipher keyed on $k^{\text{prg}}$. The 128-bit Initialization Vector (IV) carries the round index $r$ in its leading (high-order) bytes, zero-padded, while the low-order 32 bits serve as a block counter that increments for each 16-byte keystream block; as $D < 2^{32}$, the counter never overflows into the nonce, so distinct rounds yield independent keystreams. Because \textsc{CHRONOS} assumes an honest-but-curious aggregation server (Section~\ref{sec:design:threat}), we safely omit the active-adversary self-masking step present in full SecAgg, thereby reducing local computation.
To enforce execution freshness while accommodating clients that may have skipped previous rounds, the TA accepts the global round index $r$ from the untrusted Normal World but strictly enforces the invariant $r > C$. Upon validation, the TA updates its hardware-backed monotonic counter $C \leftarrow r$ and derives the mask. This ensures that a dropped client can seamlessly rejoin in a later round without counter de-synchronization, while the monotonic progression prevents mask reuse. While a malicious Normal World could intentionally advance $r$ to its maximum value to cause a local Denial of Service (DoS), this affects only availability and cannot compromise gradient confidentiality.
Client $i$ computes:
\[
  \tilde{g}_i = g_i + m_i(r) \pmod{p}
\]
and transmits $\tilde{g}_i$ to the server in a single message. \\

\textbf{Server-side aggregation (full participation).}
When all $N$ clients respond, the server computes:
\[
  \tilde{G} = \sum_{i=1}^{N} \tilde{g}_i
           = \sum_{i=1}^{N} g_i + \underbrace{\sum_{i=1}^{N} m_i(r)}_{= \mathbf{0}} = G.
\]
The mask sum is zero by the telescoping cancellation property of
Equation~\eqref{eq:mask}, requiring no de-masking step. \\

\textbf{Communication cost.}
The active-phase cost per client per round is $D \cdot \lceil \log_2 p \rceil / 8$ bytes, which is identical to plaintext FedAvg.
There is no per-round key-agreement traffic.

\subsection{Dropout Recovery via Shamir Reconstruction}
\label{sec:design:dropout}

While \textsc{CHRONOS} utilizes standard Shamir Secret Sharing (SSS) to guarantee cryptographic equivalence with established secure aggregation paradigms~\cite{SECAGG}, its architectural advantage manifests through the temporal displacement of this mechanism. Rather than executing the computationally demanding, $O(N^2)$ threshold share generation and distribution contemporaneously with the model training phase, \textsc{CHRONOS} delegates this overhead entirely to the asynchronous idle period. This decoupling isolates the critical active-phase latency from the mathematical complexity of the recovery scheme.

Table~\ref{tab:recovery_comparison} situates this design among existing dropout-recovery mechanisms. SecAgg~\cite{SECAGG} provisions recovery shares synchronously in every round and, upon dropout, requires survivors to submit shares for all $N$ clients in an additional unmasking round; its self-mask shields late-arriving gradients at the cost of this per-round interaction. MicroSecAgg~\cite{MICROSECAGG} amortizes share distribution into a one-time setup, yet its recovery remains embedded within the synchronous online protocol. Hyb-Agg~\cite{ONESHOTAGG} provides no mid-round recovery at all; a round affected by dropout must be repeated. \textsc{CHRONOS} occupies a distinct point in this design space: recovery material is provisioned once per epoch during the idle phase, so a dropout costs the active phase only a single share-collection round of $(N-k) \times k \times 32$ bytes, while the hardware-backed counter additionally protects the recovery state against rollback, a property none of the prior schemes provides. The disclosure of $\mathit{sk}_i$ upon recovery, bounded by epoch-level key rotation, is the price of this minimal active-phase footprint (Section~\ref{sec:sec:selfmask}).

\begin{table*}[t]
\centering
\caption{Comparison of dropout-recovery mechanisms in secure aggregation. ``Share distribution'' indicates when the recovery material is provisioned; ``Active-phase cost on dropout'' counts the additional synchronous rounds imposed on the latency-critical training phase. For \textsc{CHRONOS}, $k$ denotes the number of dropped clients.}
\label{tab:recovery_comparison}
\small
\begin{tabular}{lllll}
\toprule
\textbf{Scheme} & \textbf{Share distribution} & \textbf{Active-phase cost on dropout} & \textbf{Disclosure upon recovery} & \textbf{Rollback prot.} \\
\midrule
SecAgg~\cite{SECAGG} & Every round (synchronous) & One unmasking round; survivors & $\mathit{sk}_i$ of dropped clients; & None \\
 & & submit shares for all $N$ clients & self-mask shields late arrivals & \\
MicroSecAgg~\cite{MICROSECAGG} & One-time setup & Reconstruction messages within & Round-specific masking & None \\
 & & the online protocol & material only & \\
Hyb-Agg~\cite{ONESHOTAGG} & One-time (ECDH masking) & No mid-round recovery; & N/A & None \\
 & & round must be repeated & & \\
\textbf{\textsc{CHRONOS}} & Once per epoch, & One share-collection round, & $\mathit{sk}_i$ for remainder of epoch; & Hardware \\
 & during idle phase & $(N-k) \times k \times 32$ bytes total & bounded by key rotation & (RPMB) \\
\bottomrule
\end{tabular}
\end{table*}

\textbf{Mid-round dropout.}
Suppose $k$ clients drop out after the training round begins, failing to transmit
their masked gradients. To ensure successful recovery, the number of surviving clients must be sufficient to meet the Shamir reconstruction threshold ($N - k \ge t$).
Let $\mathcal{D} \subset [N]$ be the set of dropped clients and
$\mathcal{A} = [N] \setminus \mathcal{D}$ the set of survivors.
The server receives $\{\tilde{g}_j\}_{j \in \mathcal{A}}$ and must recover the aggregate
$G = \sum_i g_i$.
It observes:
\[
  \sum_{j \in \mathcal{A}} \tilde{g}_j
  = \sum_{j \in \mathcal{A}} g_j
  + \sum_{j \in \mathcal{A}} m_j(r)
  = \sum_{j \in \mathcal{A}} g_j
  - \sum_{i \in \mathcal{D}} m_i(r),
\]
where the last equality follows from the full cancellation identity
$\sum_{i=1}^{N} m_i(r) = \mathbf{0}$.
The server must recover $\sum_{i \in \mathcal{D}} m_i(r)$ to correct the aggregate. \\

\textbf{Recovery protocol.}
For each dropped client $i \in \mathcal{D}$, the server broadcasts a recovery request.
Each surviving client $j \in \mathcal{A}$ invokes its TA via \textsc{DecryptShare}$(i, E_{k^{\text{enc}}_{ji}}(\sigma_{i \to j}))$ to decrypt the ciphertext received from client $i$ during the idle phase, and responds with the plaintext share $\sigma_{i \to j}$.
Once $t$ shares are collected, the server applies Shamir reconstruction to recover
$\mathit{sk}_i$, re-derives the pairwise keys via HKDF using the peer public keys already known from the idle phase, and evaluates:
\[
  m_i(r) = \sum_{j > i} \mathrm{PRG}(k^{\text{prg}}_{ij}, r) - \sum_{j < i} \mathrm{PRG}(k^{\text{prg}}_{ji}, r)
  \pmod{p}.
\]
Here, the server uses the global training round index $r$ as the PRG nonce, which directly mirrors the dropped client's intended hardware counter value.
It repeats this for every $i \in \mathcal{D}$ and corrects the aggregate:
\[
  \hat{G} = \sum_{j \in \mathcal{A}} \tilde{g}_j + \sum_{i \in \mathcal{D}} m_i(r)
  = \sum_{j \in \mathcal{A}} g_j.
\]
The result $\hat{G}$ is the correct aggregate over the surviving clients' gradients.
FedAvg weighting is then applied using the participating clients' sample counts. \\

\textbf{Recovery payload.}
Each share $\sigma_{i \to j}$ is a 32-byte element in byte-wise GF($2^8$).
Each surviving client transmits $k \times 32$ bytes (one share per dropped client). Thus, the aggregate recovery communication across all $N-k$ survivors for $k$ dropouts is $(N-k) \times k \times 32$ bytes, which is negligible relative to any gradient vector. \\

\textbf{Threshold and security.}
The recovery protocol reveals $\mathit{sk}_i$ to the server, allowing it to derive $\mathcal{S}_i$.
This is an explicit operational trade-off acceptable under our honest-but-curious threat model, as the server already coordinates the aggregation.
Once $\mathit{sk}_i$ is reconstructed, the server can compute any mask $m_i(r')$ generated by that specific client for the current epoch.
To strictly limit this exposure window, \textsc{CHRONOS} enforces periodic key rotation (e.g., at the start of each training epoch), ensuring that
recovering a key for round $r$ does not compromise gradient confidentiality from previous
epochs.
The $t$-of-$(N{-}1)$ Shamir guarantee ensures that any $t-1$ or fewer clients
colluding with the server learn nothing about $\mathit{sk}_i$. Conversely, the system tolerates at most $t-1$ colluding clients; beyond this threshold (i.e., any $t$ clients colluding with the server), the adversary can unilaterally recover any client's ephemeral private key regardless of dropout status. The threshold parameter $t$ is set at deployment time to balance dropout robustness against collusion resistance. A higher $t$ increases the number of colluding clients the system can withstand, while a lower $t$ increases the number of mid-round dropouts ($N-t$) the system can successfully recover from. Deployments with different robustness/security trade-offs can tune $t$ accordingly. \\

\textbf{Pre-establishment dropout.}
If a client has not completed idle-phase key establishment, it possesses no seeds and
peers hold no recovery shares for it.
Such a client is excluded from the cohort until it completes the idle phase.
The server defers the start of a training epoch until a sufficient quorum of
established clients is available.
This is a once-per-epoch initialisation cost, not a per-round overhead.

\section{Security Analysis}
\label{sec:security}

In this section, we analyse \textsc{CHRONOS} against the four goals stated in Section~\ref{sec:design:threat}. All results hold under the honest-but-curious server model with up to $t-1$ colluding clients.

\subsection{OS-Level Compromise Resistance}
\label{sec:sec:root}

A central security differentiator of \textsc{CHRONOS} is its resilience against a compromised operating system on the IoT device itself. We define an adversary $\mathcal{A}_{SW}$ with full root access to the client device's Normal World (Rich Execution Environment).
     In a purely software-based secure aggregation framework, $\mathcal{A}_{SW}$
     can inspect all memory regions, extracting the pairwise derived keys
     $\mathcal{K}^{\text{prg}}_i$ and the ephemeral private key $\mathit{sk}_i$.
     Once extracted, the adversary can de-mask all future gradient updates from
     that client,  breaking confidentiality.
   
    In contrast, \textsc{CHRONOS} enforces a \emph{hardware-mediated separation
     of concerns}. The ephemeral private key $\mathit{sk}_i$ is generated and
     erased entirely within the Secure World, and the resulting Diffie-Hellman shared secrets
     $\mathcal{S}_i$ are sealed in Secure Storage under the device's
     Hardware Unique Key (HUK). Under the standard TrustZone isolation
     assumption, $\mathcal{A}_{SW}$ is physically barred from accessing Secure
     World memory or Secure Storage. For instance, rather than using a standard interconnect-level TrustZone Address Space Controller (TZASC), the Rockchip RK3399 SoC enforces secure-DRAM partitioning via its proprietary Secure General Register Files (SGRF) acting as a register-based DDR region controller \cite{RK3399_TFA, RK3399_TRM}. Consequently, provided the hardware isolation holds, even a compromised OS cannot extract the keys required to de-mask gradients, preserving computational confidentiality across all training rounds.

\subsection{Replay-Resilient Freshness}
\label{sec:sec:fresh}
To prevent mask reuse and replay attacks, \textsc{CHRONOS} enforces execution freshness via a hardware-backed round counter $C$. 
The mask $m_i(r)$ is deterministically generated from the derived PRG keys
     $\mathcal{K}^{\text{prg}}_i$ and the round index $r$, which the TA binds to
     its hardware counter $C$.

In software-only implementations, $\mathcal{A}_{SW}$ can rewind the round counter to a previous value $r'$, forcing the client to reuse mask $m_i(r')$. If the same mask is applied to two different gradients $g_i(r)$ and $g_i(r')$, the adversary can compute:
\[
  \tilde{g}_i(r) - \tilde{g}_i(r') = (g_i(r) + m_i(r')) - (g_i(r') + m_i(r')) = g_i(r) - g_i(r'),
\]
leaking the change in local gradients between rounds. The architectural guarantee of execution freshness fundamentally relies on a hardware-backed monotonic counter (e.g., Replay Protected Memory Block or eFuse registers) to prevent storage rollbacks. Our Rock Pi 4 evaluation testbed utilizes an eMMC-backed Replay Protected Memory Block (RPMB) to physically enforce this monotonic invariant. To guarantee absolute, unconditional freshness, the $\textsc{GenerateMask}$ command synchronously flushes the incremented counter $C$ to persistent RPMB storage before returning the mask to the Normal World. Because $C$ is maintained exclusively within the TrustZone Secure World and backed by the eMMC controller's internal monotonic hardware, $\mathcal{A}_{SW}$ cannot rewind $C$ via software means, nor can they bypass the increment via unexpected power-cycles or kernel panics. Because OP-TEE routes RPMB requests through the Normal World \texttt{tee-supplicant}, a root-compromised OS could refuse to proxy the write. However, this only results in a local Denial of Service (DoS) by blocking mask generation; it cannot forge a successful RPMB HMAC or rollback the counter. Thus, whenever a mask is successfully generated, it is guaranteed fresh, fully satisfying the confidentiality requirement even under root compromise.

\subsection{Functional Correctness and Dropout Recovery}
\label{sec:sec:correct}
The primary functional goal is that the server accurately recovers the sum of the unmasked gradients. By construction (Equation~\ref{eq:mask}), each mask $m_i(r)$ is a sum of PRG outputs with alternating signs across all ordered pairs $(i,j)$. Summing over all clients yields:
\[
     \sum_{i=1}^{N} m_i(r) = \sum_{(i,j)\in\mathcal{P}}
     \mathrm{PRG}(k^{\text{prg}}_{ij}, r) - \sum_{(i,j)\in\mathcal{P}}
     \mathrm{PRG}(k^{\text{prg}}_{ij}, r) = \mathbf{0} \pmod{p}.
    \]
    Thus, when all clients participate, the masked gradients sum exactly to the
     plaintext aggregate: $\sum_{i=1}^{N} \tilde{g}_i = \sum_{i=1}^{N} g_i +
     \mathbf{0} = G$.
     
When $k$ clients drop out (where $k \le N-t$), the zero-sum property of the
     masks is broken. However, the server can request Shamir shares from the
     surviving clients. With $t$ shares, the server reconstructs the dropped
     client's ephemeral private key $\mathit{sk}_i$, re-derives their pairwise
     keys via HKDF using the public keys, and computes the missing mask $m_i(r)$
     using the target round index $r$ as the PRG nonce. By adding the dropped
     masks to the sum of the received masked gradients, the server reconstructs the aggregate of the surviving clients' gradients without
     requiring a new training round.

\textbf{Scope note on intra-epoch recovery leakage.} The dropout-recovery protocol reveals $\mathit{sk}_i$ to the server for the remainder of the current epoch. If client $i$ drops in round $r$ and later attempts to rejoin in round $r' > r$ within the same epoch, the server, having already reconstructed $\mathit{sk}_i$, could theoretically unmask $g_i(r')$. To prevent this, \textsc{CHRONOS} enforces a strict protocol rule: the server discards any message arriving from a client after its sender has been declared dropped within the same epoch. The client is explicitly excluded from participation until the next epoch-boundary key rotation establishes fresh ephemeral keys. This rule is enforceable under our honest-but-curious threat model, as the server executes the protocol faithfully.

\subsection{Analysis of the Self-Masking Trade-off}
\label{sec:sec:selfmask}
A recognized vulnerability within federated secure aggregation architectures materializes when a central aggregator either falsely declares a client $i$ as dropped or receives a delayed message out of band. Because the server recovers the ephemeral private key $\mathit{sk}_i$ through the Shamir dropout protocol to cancel the missing client's pairwise masks, it possesses the requisite cryptographic material to de-mask a late-arriving gradient $g_i$. Standard implementations of SecAgg~\cite{SECAGG} mitigate this vulnerability by requiring each client to superimpose a secondary, independent "self-mask" derived from a distinct secret seed $b_i$. Because reconstructing $\mathit{sk}_i$ does not reveal $b_i$, the secondary mask preserves privacy during false-dropout conditions.

\textsc{CHRONOS} explicitly omits this self-mask, a deliberate architectural decision prioritizing the strict latency constraints of IoT environments over protection against an actively malicious aggregator. Retaining the self-mask mandates a secondary interactive round during the active phase, forcing surviving clients to negotiate and transmit the aggregated cancellation values for the dropped clients' self-masks. Such multi-round synchronization incurs substantial blocking overhead, nullifying the fundamental single-message active-phase latency advantage that renders \textsc{CHRONOS} scalable.

This trade-off is strictly bound by the honest-but-curious threat model established in Section~\ref{sec:design:threat}. We assert that because the aggregator executes the protocol faithfully, it does not maliciously simulate dropouts to harvest keys. For instances of genuine network delay wherein a masked gradient arrives after a dropout recovery has been initiated, the protocol strictly requires the server to discard the late-arriving packet without attempting decryption. Consequently, \textsc{CHRONOS} exchanges defense against an actively malicious false-dropout vector for a mathematically strict $O(N)$ active phase, an optimization deemed essential for deploying federated learning across energy-constrained sensor networks.

\subsection{Side-Channel Resilience}
\label{sec:sec:sidechannel}
Within the established federated learning threat model, the adversary is presumed to operate remotely, executing system-level intrusions via the network interface. Consequently, physical hardware attacks requiring direct, on-site access to the silicon (e.g., macroscopic probing, physical fault injection, or decapping) are fundamentally out of scope. However, while ARM TrustZone provides strong architectural isolation between the Normal and Secure Worlds, recent research has demonstrated vulnerabilities that exploit shared microarchitectural resources accessible via software. Specifically, attacks such as CLKscrew~\cite{CLKSCREW} and VoltJockey~\cite{VOLTJOCKEY} leverage the shared Dynamic Voltage and Frequency Scaling (DVFS) and power management interfaces to induce faults or observe timing variations in Secure World execution from a compromised Normal World OS.

\textsc{CHRONOS} is designed with three inherent mitigations against such side-channel and fault-injection threats.

The system minimizes the \emph{window of vulnerability}: unlike TEE-based FL frameworks that repeatedly load large parameter chunks into the enclave during training~\cite{PPFL,GRADSEC}, \textsc{CHRONOS} restricts Secure World execution exclusively to a single mask generation call per round. As shown in Section~\ref{sec:eval:rq5}, this execution window is extremely short (e.g., 44.5\,ms for the small CNN), significantly increasing the difficulty for an adversary to synchronize and mount high-resolution side-channel attacks.

\textsc{CHRONOS} relies on constant-time cryptographic primitives. The TA leverages ARMv8 Crypto Extensions for hardware-accelerated AES-GCM operations (used in sealing), while the AES-128-CTR hardware instructions used for mask generation are inherently constant-time. This ensures the TEE's core timing profile is independent of secret seed values. While the subsequent rejection sampling step required for field reduction exhibits data-dependent timing, this variation reveals only statistical properties of a pseudorandom output and does not provide a practical extraction path for the underlying PRG seeds.

The integrity of the sealed keys is protected via AES-256-GCM. Any attempt by an adversary to induce bit-flips in Secure Storage through voltage manipulation (e.g., undervolting during a memory write) will be detected as an authentication failure during the subsequent unsealing process. This combination of a minimal TEE footprint and authenticated sealing substantially increases the cost of DVFS-based fault injection. We further observe that, within the per-round Secure World residency, the fixed RPMB flush ($\approx$39\,ms) dominates, while the only secret-dependent computation, namely $\mathbb{F}_p$ rejection sampling over pseudorandom output, occupies a sub-6\,ms window and reveals merely statistical properties of the keystream rather than seed material. This further narrows the temporal surface an adversary could synchronize against for high-resolution timing or voltage-glitch attacks. A formal physical side-channel and fault-injection campaign, which requires instrumented silicon and controlled glitching equipment, remains out of scope for this architectural paper and is identified as future work.

\subsection{Defense-in-Depth and Mitigation Strategy}
\label{sec:sec:idle}
If a device has not completed the idle-phase key establishment, it possesses no derived keys and cannot generate masks. In this case, \textsc{CHRONOS} degrades gracefully: the client is excluded from the current training epoch rather than transmitting an unprotected gradient. This resilience ensures that devices only participate when cryptographic protections are fully established. 

Finally, while \textsc{CHRONOS} protects against gradient inversion, it does not inherently prevent membership inference attacks against the aggregated model. Furthermore, we assume an honest-but-curious server that computes the aggregate faithfully. A fully malicious server could attempt to exclude updates or return corrupted models. As discussed in Section~\ref{sec:design:idle}, MITM attacks during key exchange are already mitigated via certificate-based Authenticated Key Exchange during device provisioning. To defend the active phase against malicious aggregation, future work might extend the system with zero-knowledge proofs. This combination of hardware isolation, monotonic counters, and authenticated provisioning provides a defense-in-depth architecture suitable for high-stakes IoT federated learning.

\section{Implementation}
\label{sec:impl}

\textsc{CHRONOS} is implemented in two components: a Trusted Application (TA) that executes in the ARM TrustZone Secure World, and a host-side suite of Normal World programs that drive the idle-phase key establishment and the active-phase gradient masking.

\subsection{Trusted Application}
\label{sec:impl:ta}

Specifically, the TA is written in C using the OP-TEE SDK and compiled for the ARM Cortex-A72
Rockchip RK3399 SoC (2$\times$ Cortex-A72 + 4$\times$ Cortex-A53 big.LITTLE) of the Rock Pi 4, targeting OP-TEE 4.4.0.
The same TA source is additionally compiled, without modification, for the Rockchip RK3588S SoC (4$\times$ Cortex-A76 + 4$\times$ Cortex-A55) of the Orange Pi 5 under the identical OP-TEE 4.4.0 configuration, for both SoCs expose the GlobalPlatform Internal Core API and ARMv8 Crypto Extensions required by the TA.
It implements six GlobalPlatform TEE Internal Core API commands:

\begin{itemize}
\item \textbf{\texttt{KEYGEN}}: invokes the GlobalPlatform \texttt{TEE\_GenerateKey} API for the elliptic curve, retains the private key
  $\mathit{sk}_i$ in a Secure World object handle, and writes only the public key
  $\mathit{pk}_i$ (32 bytes) to a Normal World shared memory buffer.
  Since $\mathit{sk}_i$ never leaves the Secure World, a compromised host OS cannot
  extract it irrespective of timing.

\item \textbf{\texttt{COMPUTE\_SEEDS}}: accepts a shared memory buffer
  containing $N-1$ peer public keys $\{pk_j\}_{j \neq i}$, invokes
  \texttt{TEE\_DeriveKey} (using the \texttt{TEE\_ALG\_X25519} identifier, requiring \texttt{CFG\_CRYPTO\_X25519=y}) for each peer using the retained
  $\mathit{sk}_i$, computes the raw Diffie-Hellman shared secrets, applies HKDF to derive the immediate encryption keys for share distribution, stores the $N-1$ raw secrets in a Secure World memory
  buffer, and retains $\mathit{sk}_i$ in the key object handle for use in the subsequent
  SEAL command.
  Both the derived keys and $\mathit{sk}_i$ exist only within the Secure World at this point.

\item \textbf{\texttt{SEAL}$(t)$}: performs four actions atomically.
  First, it generates Shamir secret shares of the ephemeral private key $\mathit{sk}_i$
  using a $t$-of-$(N{-}1)$ scheme over byte-wise GF($2^8$). It encrypts each share $\sigma_{i \to j}$ (32 bytes) under the corresponding derived encryption key $k^{\text{enc}}_{ij}$ using AES-128-GCM and freshly generated IV and writes the ciphertexts to a Normal World shared memory buffer for distribution to peers.
  Second, it encrypts the raw Diffie-Hellman shared secret set $\mathcal{S}_i$ under AES-256-GCM with a key derived from the
  device Hardware Unique Key (HUK), and writes the ciphertext and 16-byte authentication
  tag to OP-TEE Secure Storage atomically.
  Third, it initialises the round counter $C \leftarrow 0$.
  Fourth, $\mathit{sk}_i$ is securely erased from Secure World memory.
  Total Secure Storage usage: $(N-1) \times 32 + 16_{\text{tag}} + 4_{\text{counter}}$
  bytes; for $N = 32$, the theoretical total is 1012\,bytes (1016\,bytes as measured).

\item \textbf{\texttt{GENERATE\_MASK}$(D, r)$}: decrypts and authenticates the
  shared secret set from Secure Storage, derives the PRG keys via HKDF, validates $r > C$, evaluates Equation~\eqref{eq:mask} using AES-128-CTR as the
  PRG (seeded with $k^{\text{prg}}_{ij}$, nonce $= r$), atomically updates $C \leftarrow r$ in Secure RAM (synchronously flushing to the eMMC RPMB), and returns the
  $D$-dimensional mask vector to the Normal World via a shared memory buffer.
  Returns \texttt{ERR\_NOT\_READY} if no key set has been sealed.

\item \textbf{\texttt{DECRYPT\_SHARE}$(i, ct)$}: accepts a ciphertext of a Shamir share from dropped client $i$, decrypts the shared secret set from Secure Storage, derives the appropriate encryption key $k^{\text{enc}}_{ji}$ via HKDF, decrypts the share using AES-128-GCM, and returns the plaintext share $\sigma_{i \to j}$ to the Normal World.

\item \textbf{\texttt{PEEK\_COUNTER}}: returns the current value of $C$
  without modifying any state, enabling the Normal World daemon to verify round
  synchronisation.
\end{itemize}

We employ AES-128-CTR as the PRG. By utilizing the RK3399's ARMv8 Crypto Extensions through the stock \texttt{mbedTLS} ARMv8-CE path configured in OP-TEE 4.4.0, the TA achieves a stable hardware-accelerated throughput of $\approx 1.26$\,GB/s. Because $p = 2^{31}-1$ is a Mersenne prime, the TA masks each 32-bit chunk with \texttt{0x7FFFFFFF} before field reduction, which lowers the rejection-sampling rate to $1/2^{31}$ rather than removing the rejection step itself (the data-dependent branch is retained; see Section~\ref{sec:sec:sidechannel}). This optimization halves the keystream volume compared to naive 32-bit rejection, allowing the TA to compute the heavy $O(N)$ mask generation entirely within the sub-100\,ms active window.

The TA binary size is 48 KB (compiled with \texttt{-Os}).
The Secure World runtime memory footprint (stack, heap, session context) is 18 KB.

\subsection{Idle-Phase Daemon}
\label{sec:impl:daemon}

The idle-phase daemon is a Python program (approximately 280 lines of code) that
monitors device operational state and orchestrates the once-per-epoch key establishment and
share distribution.

When the device transitions to the idle state, the daemon executes the following steps:
\begin{enumerate}
\item Invokes \texttt{KEYGEN} to generate the ephemeral keypair entirely
  inside the TA, receiving only $\mathit{pk}_i$.
\item Sends $\mathit{pk}_i$ to the server and receives the peer public keys
  $\{\mathit{pk}_j : j \neq i\}$ relayed by the server.
\item Invokes \texttt{COMPUTE\_SEEDS} with the received peer keys;
  the TA computes all DH secrets and retains $\mathit{sk}_i$ in Secure World memory
  for use in the subsequent SEAL command.
\item Invokes \texttt{SEAL}$(t)$, receiving the $N-1$ ciphertexts
  $\{E_{k^{\text{enc}}_{ij}}(\sigma_{i \to j})\}_{j \neq i}$ in return; the shared secrets are now sealed in Secure Storage.
\item Distributes the ciphertext $E_{k^{\text{enc}}_{ij}}(\sigma_{i \to j})$ to peer $j$ for each $j \neq i$, relayed
  by the server.
\item Receives and stores the ciphertext $E_{k^{\text{enc}}_{ji}}(\sigma_{j \to i})$ from each peer $j \neq i$ in local
  (non-TEE) storage.
\end{enumerate}

The once-per-epoch idle-phase cost is dominated by the $N-1$ Curve25519 DH operations and the
key relay round trip.
On Rock Pi 4 hardware, the 31 DH operations take approximately 280 ms of local compute. While hand-optimized AArch64 X25519 assembly can execute in microseconds, this duration reflects OP-TEE's default C-based \texttt{mbedTLS} implementation, the sequential TEE boundary crossings, and the subsequent HKDF derivations. Including network round-trips for public-key and share relay, the end-to-end idle-phase time is $\approx$ 350 ms.
Once complete, the idle daemon has no further work until the next key-rotation epoch.

\subsection{Active-Phase Training Wrapper}
\label{sec:impl:wrapper}

The active-phase training wrapper (approximately 180 lines of Python code) subclasses
the Flower \texttt{NumPyClient} class, overriding \texttt{fit} to inject the masking step.

When \texttt{fit} is called, the wrapper:
\begin{enumerate}
\item Invokes \texttt{PEEK\_COUNTER} to confirm key establishment is complete.
\item If \texttt{ERR\_NOT\_READY} is returned, raises a
  \texttt{DropoutException} and skips the round.
\item Computes the local gradient $g_i$ via PyTorch as a vector of 32-bit floating-point values (FP32).
\item \textbf{Gradient Quantization:} To perform cryptographic masking over a finite field, the wrapper scales the FP32 gradients by a fixed factor $S = 2^{16}$ and rounds them to integers. These integers are mapped to positive elements in $\mathbb{F}_p$ by shifting the representation domain to $[0, p-1]$. We use $p = 2^{31} - 1$ (Mersenne prime), which accommodates aggregates up to $N \times S \times \max\|g_i\|_\infty \approx 32 \times 2^{16} \times 1000 \approx 2.09 \times 10^9$ without aliasing (wraparound) in $\mathbb{F}_p$ for the evaluated $N$ and $D$. The constant domain shift is deterministically subtracted by the server during aggregation. We empirically verified $\max \|g_i\|_\infty < 1000$ across all training runs; gradients exceeding this bound would be clipped to maintain field-arithmetic correctness.
\item Invokes \texttt{GENERATE\_MASK}$(D, r)$ in a single TEE call,
  receiving the pseudo-random mask vector $m_i(r) \in \mathbb{F}_p^D$. The TA generates this vector by extracting 32-bit chunks from the AES-128-CTR keystream and reducing them modulo $p$.
\item Computes $\tilde{g}_i = (g_i + m_i(r)) \pmod{p}$ element-wise via NumPy.
\item Returns $\tilde{g}_i$ to the Flower communication layer.
\end{enumerate}

The single TEE call per round crosses the Normal-World/Secure-World boundary exactly once
regardless of model dimension. Upon receiving the masked gradients, the server sums them modulo $p$, recovers the integer aggregate, maps it back to the signed domain, and scales by $S^{-1}$ to restore the FP32 aggregate gradient.

\subsection{Server Aggregation and Dropout Recovery}
\label{sec:impl:server}

The server aggregation component (approximately 500 lines of Python code) subclasses
the Flower \texttt{Strategy} class and overrides \texttt{aggregate\_fit}.

\textbf{Normal-round aggregation.}
When all $N$ clients respond, the server computes $\tilde{G} = \sum_i \tilde{g}_i = G$
as explained in Section~\ref{sec:sec:correct}.
No de-masking step is required; the pairwise cancellation is exact by construction.

\textbf{Dropout recovery.}
When the server detects that $k \le N-t$ clients have failed to respond within a timeout,
it executes the recovery protocol of Section~\ref{sec:design:dropout}.
For each dropped client $i \in \mathcal{D}$, the server broadcasts a
\textsc{RecoveryRequest}$(i)$ message to all surviving clients.
Each surviving client $j \in \mathcal{A}$ invokes \texttt{DECRYPT\_SHARE} to decrypt their locally stored ciphertext and responds with the plaintext share $\sigma_{i \to j}$ (32 bytes) via the Flower \texttt{evaluate} channel.
The server collects $t$ shares, applies Lagrange interpolation over byte-wise GF($2^8$) to
reconstruct $\mathit{sk}_i$, re-derives the pairwise keys via HKDF
from the peer public keys already known from the idle phase, evaluates $m_i(r)$, and adds
it to the partial sum.
FedAvg weighting is applied using the sample counts of the participating clients in
$\mathcal{A}$, and the updated global model $w^{(r+1)}$ is broadcast as usual.

\section{Evaluation}
\label{sec:eval}

We conduct an extensive evaluation of \textsc{CHRONOS} to answer six research questions:

\begin{itemize}
\item \textbf{RQ1 (Active-Phase Latency and Energy):} Does \textsc{CHRONOS} reduce per-round active-phase aggregation latency and energy consumption
        compared to SecAgg and other baselines?

\item \textbf{RQ2 (Storage Overhead):} What is the Secure World storage footprint of
\textsc{CHRONOS}, and how does it scale with model dimension and training horizon?

\item \textbf{RQ3 (Scalability):} How does \textsc{CHRONOS}'s per-round latency scale
as the number of participating clients $N$ increases?

\item \textbf{RQ4 (Dropout Recovery):} Does \textsc{CHRONOS}'s Shamir-based recovery
preserve correct aggregate computation under systematic client dropout, and what is the
recovery communication overhead?

\item \textbf{RQ5 (TEE Overhead):} What fraction of active-phase round time is attributable to
the TrustZone context-switch for \texttt{GENERATE\_MASK}?

\item \textbf{RQ6 (Empirical Privacy):} Does \textsc{CHRONOS} successfully thwart state-of-the-art optimization-based gradient inversion attacks compared to plaintext FL?
\end{itemize}

Let us first present our experimental setup, before reporting the evaluation results for the six research questions mentioned earlier. \\

\textbf{Hardware.} 
To evaluate performance under realistic system heterogeneity, we deploy a 32-node physical testbed comprising two distinct classes of Cortex-A edge gateways. The primary cluster consists of 20 Rock Pi 4 devices (Model B; 4\,GB RAM, RK3399 SoC with Cortex-A72/A53). The secondary cluster comprises 12 Orange Pi 5 devices (8\,GB RAM, RK3588S SoC with Cortex-A76/A55). Both hardware profiles natively feature ARMv8 Crypto Extensions and hardware-enforced secure-DRAM partitioning, realized on Rockchip SoCs via register-based DDR region controllers rather than a standard TZASC (see Section~\ref{sec:sec:root}). Each device is equipped with an eMMC module to enable RPMB secure storage, running OP-TEE 4.4.0 built with \texttt{CFG\_TZDRAM\_SIZE=32M} (Secure World memory allocation) and \texttt{CFG\_WITH\_VFP=y} (enabling hardware-accelerated AES-128-CTR). The aggregation server is a Google Cloud Platform \texttt{c2-standard-8} instance (8 vCPUs, 32\,GB RAM). Clients communicate over a local area network ($<$1\,ms round-trip). \\

\textbf{Energy measurement.} 
To isolate the cryptographic energy overhead during the high-speed active phase, we designed a dedicated hardware profiling rig. Because the active phase duration (75--450\,ms) is shorter than the reliable integration window of standard USB-C power monitors, we instrumented the Rock Pi 4's 5\,V input rail with a 10\,m$\Omega$ precision shunt resistor and a Texas Instruments INA226 16-bit power monitor. The monitor was polled out-of-band via I$^2$C at 400\,kHz by a bare-metal microcontroller, yielding a $\sim$900\,Hz sampling rate. This high-resolution logging was synchronized to the active-phase boundaries via hardware GPIO toggles, allowing us to perform clean trapezoidal integration ($\int V \times I \, dt$) over the entire active training window without being affected by OS scheduler jitter. We report measurements only for the small CNN, as the medium CNN's energy consumption is dominated by transmission overhead identically across all systems. \\

\textbf{Datasets.}
We evaluate on CIFAR-10 (50,000 training images, 10 classes) with Dirichlet non-IID
partitioning ($\alpha = 0.5$), FEMNIST, and the UCI Human Activity Recognition (HAR) dataset.
To accommodate the expanded heterogeneous cohort size ($N=32$) while maintaining meaningful intra-client
datasets, we filter FEMNIST to the 10 digit classes and partition it naturally
by writer identity. We acknowledge this is a non-standard filtering that affects direct comparability of raw accuracy numbers with prior work evaluating on the full 62-class FEMNIST, but it preserves the non-IID characteristics necessary for evaluating dropout robustness. For UCI-HAR, which contains 561 pre-extracted features from smartphone
accelerometer and gyroscope sensors across 6 activities, we partition the data naturally
by assigning 1-2 subjects per client, a configuration highly representative of realistic IoT
sensing data distributions. \\

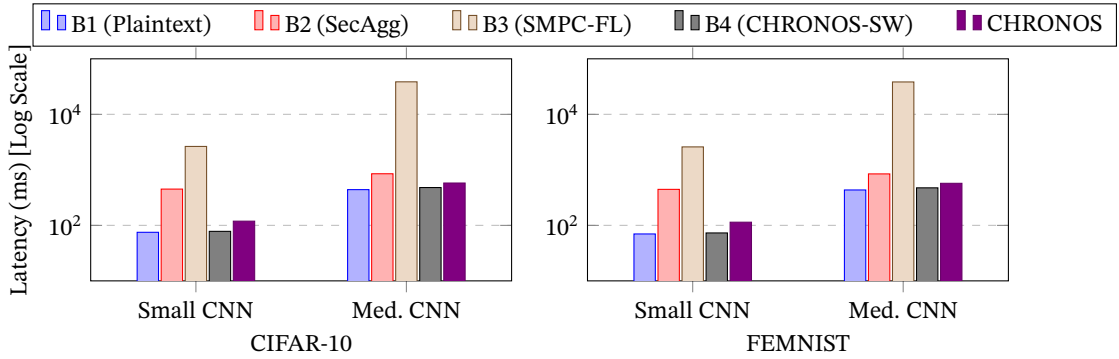
\begin{figure*}[t]
\centering
\begin{tikzpicture}

\begin{axis}[
 name=plot1, 
 ybar=1pt,
 ymode=log,
 width=0.4\textwidth,
 height=4.5cm,
 enlarge x limits=0.5,
 ylabel={Latency (ms) [Log Scale]},
 symbolic x coords={Small CNN, Med. CNN},
 xtick=data,
 bar width=8pt,
 xlabel={CIFAR-10},
 ymin=10, ymax=100000,
 ymajorgrids=true,
 grid style=dashed,
 legend style={at={(1.15, 1.25)}, anchor=north, legend columns=6, /tikz/every even column/.append style={column sep=0.5cm}},
 legend entries={B1 (Plaintext), B2 (SecAgg), B3 (SMPC-FL), B4 (CHRONOS-SW), \textsc{CHRONOS}},
]
\addplot coordinates {(Small CNN, 75) (Med. CNN, 440)};
\addplot coordinates {(Small CNN, 450) (Med. CNN, 850)};
\addplot coordinates {(Small CNN, 2640) (Med. CNN, 38500)};
\addplot coordinates {(Small CNN, 78) (Med. CNN, 480)};
\addplot coordinates {(Small CNN, 119) (Med. CNN, 579)};
\end{axis}

\begin{axis}[
 name=plot2,
 at={(plot1.south east)}, 
 anchor=south west,       
 xshift=1cm,            
 ybar=1pt,
 ymode=log,
 width=0.4\textwidth,
 height=4.5cm,
 enlarge x limits=0.5,
 symbolic x coords={Small CNN, Med. CNN},
 xtick=data,
 bar width=8pt,
 xlabel={FEMNIST},
 ymin=10, ymax=100000,
 ymajorgrids=true,
 grid style=dashed
]
\addplot coordinates {(Small CNN, 70) (Med. CNN, 434)};
\addplot coordinates {(Small CNN, 445) (Med. CNN, 845)};
\addplot coordinates {(Small CNN, 2590) (Med. CNN, 38300)};
\addplot coordinates {(Small CNN, 73) (Med. CNN, 474)};
\addplot coordinates {(Small CNN, 114) (Med. CNN, 572)};
\end{axis}

\end{tikzpicture}
\caption{Per-round active-phase aggregation latency (log scale) for CIFAR-10 and FEMNIST across evaluated systems.}
\label{fig:latency_charts}
\end{figure*}

\textbf{Models.}
For the vision tasks, we use a small CNN with approximately 50,000 parameters ($D = 50{,}000$)
and a medium CNN with approximately 1,000,000 parameters ($D = 1{,}000{,}000$).
For the UCI-HAR task, we employ a lightweight 1D-CNN standard for the domain, scaling
the architecture to yield a comparable parameter footprint ($D \approx 50{,}000$). Because
the gradient dimensionality matches the vision small CNN, we report latency, energy, and
storage overheads for the vision tasks; the HAR IoT footprint is definitionally identical
to the small CNN footprint. \\

\textbf{Federated learning protocol.}
All systems use FedAvg with 50 global rounds (with epoch-boundary key rotation performed every 10 rounds), full client participation ($N = 32$), 5 local
epochs per round, and learning rate $\eta = 0.01$.
All measurements are repeated over 5 independent random seeds; we report the empirical mean
and 95\% confidence intervals.
Confidence intervals are computed assuming a normal distribution across the seeds; we acknowledge
that physical I/O and OS scheduler jitter may introduce some non-normal variance not captured
by the smoothed point estimates. \\

\textbf{Baselines.}
\begin{itemize}
\item \textbf{B1 (Plaintext FedAvg):} Standard  FedAvg with no privacy protection.
Lower bound on latency and energy; upper bound on privacy leakage.

\item \textbf{B2 (SecAgg):} Bonawitz et al.\ secure aggregation~\cite{SECAGG} with synchronous
key agreement.
This is the most direct baseline: \textsc{CHRONOS} uses the same cryptographic primitive
as SecAgg but moves the key exchange to a pre-computation phase. While MicroSecAgg~\cite{MICROSECAGG} represents a newer single-server variant with similar $O(N^2)$ key-establishment cost, we treat B2 as representative of the synchronous key-agreement paradigm.

\item \textbf{B3 (SMPC-FL):} Generic Secure Multiparty Computation (SMPC) via the MP-SPDZ framework. As discussed in Section~\ref{sec:related}, this baseline is included to empirically demonstrate the prohibitive computational penalty of using heavy, Oblivious Transfer (OT)-based protocols for simple additive aggregation tasks.

\item \textbf{B4 (CHRONOS-SW):} A software-only ablation in which DH keys and counters are stored in the Normal World (filesystem) rather than the TEE. This baseline isolates the latency and energy contribution of the "idle-window" scheduling decision from the hardware-isolation security contribution of the TEE, similar in architecture to prior software-only pre-computation designs~\cite{ONESHOTAGG}.
\end{itemize}

\textbf{Model utility.}
Because the pairwise masks cancel exactly upon aggregation (Section~\ref{sec:sec:correct}), \textsc{CHRONOS} is utility-preserving by construction. Table~\ref{tab:acc} confirms this empirically: across CIFAR-10, FEMNIST, and UCI-HAR, the final test accuracy of \textsc{CHRONOS} is identical to the plaintext FedAvg baseline (B1). We therefore devote the remainder of the evaluation to the system overheads introduced by the privacy layer rather than to model quality. \\

\begin{table*}[t]
\centering
\caption{Final test accuracy (\%) after 50 rounds (mean $\pm$ 95\% CI).
CHRONOS matches B1 exactly due to complete mask cancellation.}
\label{tab:acc}
\small
\begin{tabular}{lccccc}
\toprule
& \multicolumn{2}{c}{\textbf{CIFAR-10}} & \multicolumn{2}{c}{\textbf{FEMNIST}} & \textbf{UCI-HAR} \\
\cmidrule(lr){2-3} \cmidrule(lr){4-5} \cmidrule(lr){6-6}
\textbf{System} & \textbf{Small CNN} & \textbf{Med. CNN} & \textbf{Small CNN} & \textbf{Med. CNN} & \textbf{1D-CNN} \\
\midrule
B1 (Plaintext)   & $63.5 \pm 0.7$ & $71.2 \pm 0.6$ & $79.8 \pm 0.6$ & $82.5 \pm 0.5$ & $88.7 \pm 1.5$ \\
\textbf{CHRONOS} & $63.5 \pm 0.7$ & $71.2 \pm 0.6$ & $79.8 \pm 0.6$ & $82.5 \pm 0.5$ & $88.7 \pm 1.5$ \\
\bottomrule
\end{tabular}
\end{table*}

\subsection{RQ1: Active-Phase Latency and Energy Consumption}
\label{sec:eval:rq1}

Figure~\ref{fig:latency_charts} reports per-round active-phase aggregation latency (ms) for each system.
Active-phase aggregation latency is measured strictly as the client-side round-trip time:
  from the completion of local gradient computation in round $r$ (when cryptographic masking begins), to the receipt of
  the updated global model $w^{(r+1)}$ for the subsequent round.
Because the local PyTorch training time (tens of seconds for 5 epochs) is identical across all baselines, isolating the aggregation phase allows us to accurately quantify the precise network and cryptographic overhead introduced by the privacy protocols.
For \textsc{CHRONOS}, the once-per-epoch idle-phase key establishment is excluded, as it occurs
outside the active training loop.

Our evaluation demonstrates that \textsc{CHRONOS} recovers near-plaintext latency while providing OS-level compromise resistance.
As shown in Figure~\ref{fig:latency_charts},  \textsc{CHRONOS}'s active-phase aggregation latency (119\,ms and 579\,ms
for the small and medium CNNs, respectively) introduces a small, constant overhead over the plaintext baseline B1 (75\,ms and 440\,ms).

The absolute overhead of the cryptographic protection is approximately $44$ ms and $139$ ms for the small and medium CNNs respectively, attributable solely to the TEE context-switch, mask derivation, and the RPMB flush.
Compared to the synchronous SecAgg baseline B2, \textsc{CHRONOS} achieves a 74\% reduction
in active-phase aggregation latency for the small CNN ($D=50$K).

\textbf{Quantitative Comparison with State-of-the-Art.}
Beyond the canonical SecAgg baseline, we quantitatively map \textsc{CHRONOS}'s performance against two recent state-of-the-art frameworks: MicroSecAgg~\cite{MICROSECAGG} and Hyb-Agg~\cite{ONESHOTAGG}. Because both schemes reuse the same cryptographic primitives as our existing baselines and differ primarily in their active-phase round structure, we obtain their latencies by configuring our SecAgg (B2) and decoupled (B4) harnesses to the published active-phase round topologies on the identical $N=32$ physical testbed, rather than re-implementing each system end-to-end. MicroSecAgg compresses SecAgg's four-round active phase down to three synchronous rounds; measured under this configuration on the same INA226 rig, it yields an $N=32$ active-phase latency of 315.6\,ms and a corresponding energy footprint of 1618.4\,mJ (a $1.4\times$ energy reduction over SecAgg, yet still $2.8\times$ the energy of \textsc{CHRONOS}). While an improvement over SecAgg, this still represents a $2.6\times$ latency overhead compared to \textsc{CHRONOS}'s single-message active phase. Hyb-Agg achieves a streamlined two-round active phase via one-shot ECDH masking, reaching a highly competitive 204.8\,ms latency. However, Hyb-Agg explicitly sacrifices mid-round dropout recovery to achieve this speed; if a single client drops, the entire active phase must be repeated, destroying the latency and energy profile of the cohort. \textsc{CHRONOS} outperforms both SOTA schemes, achieving the lowest active-phase latency (119\,ms) and energy footprint (576.5\,mJ) while fully preserving robust dropout recovery.

\textbf{Hardware Heterogeneity Resilience.}
Our 32-node testbed inherently captures the performance disparity between the Cortex-A72 (Rock Pi 4) and Cortex-A76 (Orange Pi 5) architectures. In synchronous protocols like B2 (SecAgg), the server must wait for all clients to finish the computationally heavy $O(N^2)$ key-agreement phase, causing the slower Cortex-A72 nodes to act as stragglers and bottleneck the entire cohort. Specifically, we observed that Cortex-A72 nodes require 31.4\% more time for DH exponentiations than the Cortex-A76 nodes. By decoupling this heavy cryptography to the asynchronous idle phase, \textsc{CHRONOS} ensures that the active-phase latency is determined solely by the lightweight AES-CTR masking (which executes in under 6\,ms for the small CNN on both architectures) and the fixed RPMB flush. Consequently, \textsc{CHRONOS} completely eliminates cryptographic straggler effects during the latency-critical training round.

\textbf{Network Diversity and WAN Emulation.}
To evaluate robustness under diverse network conditions beyond our local LAN setup ($<$1\,ms RTT), we utilized \texttt{netem} at the gateway router to inject variable WAN latencies (50\,ms, 100\,ms, and 200\,ms RTT) and 1\% packet loss. Under a 100\,ms RTT, B2 (SecAgg)'s multi-round synchronous design suffers severe degradation, with active-phase latency ballooning from 450\,ms to $1{,}214.3$\,ms due to compounded network round-trips. In contrast, \textsc{CHRONOS}'s single-message active phase absorbs the 100\,ms RTT gracefully, increasing from 119\,ms to 224.6\,ms. This confirms that phase-decoupling is exceptionally resilient to the highly variable network topologies typical of real-world IoT deployments.

The comparison with B4 (\textsc{CHRONOS-SW}) shows that the latency cost of TrustZone is modest ($\approx$41--99 ms). While \textsc{CHRONOS-SW} achieves slightly lower latency by avoiding the TEE context switch and RPMB flush, it offers no protection against a root-compromised host OS, as seeds and counters are stored in software. \textsc{CHRONOS} therefore matches the security guarantees of TEE-based designs while approaching the latency of software-only decoupled schemes, at a cost of only 41\,ms (53\%) for the small CNN and 99\,ms (21\%) for the medium CNN.

For the medium CNN ($D=1$M), the relative reduction drops to $\sim$32\% because gradient
transmission over the IoT network interface begins to dominate the per-round duration,
yet \textsc{CHRONOS} remains significantly faster than SecAgg by eliminating the
synchronous key-exchange rounds.
B3 is the slowest system by one to two orders of magnitude, confirming that OT-based
input masking is an unnecessary overhead for linear gradient aggregation. \\

\textbf{Idle-phase amortised cost.}
The idle-phase key establishment is a once-per-epoch cost. On the heterogeneous hardware, the end-to-end idle-phase setup requires approximately 350\,ms. For our evaluated small CNN, \textsc{CHRONOS} saves $\approx$331\,ms of active-phase latency per round compared to SecAgg (450\,ms vs.\ 119\,ms). With an epoch length of 10 rounds, the 350\,ms idle-phase investment yields over 3.3\,seconds of active-phase savings, representing an amortised setup overhead of less than 11\% of the active-phase time saved per epoch. \\

\textbf{Energy Footprint.} 
Table~\ref{tab:energy} reports the empirically integrated per-round active-phase energy (mJ), captured directly by the INA226 hardware monitor. We report energy for B1, B2, B4, and \textsc{CHRONOS} only; B3's energy footprint is omitted as it is three orders of magnitude higher and provides no additional insight.

\begin{table}[t]
\centering
\caption{Empirically measured per-round active-phase energy (mJ, mean $\pm$ 95\% CI, 5 seeds). Lower is better. Energy is integrated directly from the INA226 hardware power monitor traces synchronized to protocol boundaries.}
\label{tab:energy}
\small
\begin{tabular}{lcc}
\toprule
\cmidrule(lr){2-3}
\textbf{System} & \begin{tabular}{@{}c@{}}\textbf{Measured active-phase} \\ \textbf{energy (mJ)}\end{tabular} & \textbf{Reduction vs.\ B2} \\
\midrule
B1 (Plaintext) & $318.6 \pm 12.4$ & $7.1\times$ \\
B2 (SecAgg) & $2248.3 \pm 46.1$ & $1\times$ \\
B4 (CHRONOS-SW) & $334.8 \pm 14.2$ & $6.7\times$ \\
\textbf{CHRONOS} & $576.5 \pm 18.3$ & $3.9\times$ \\
\bottomrule
\end{tabular}
\end{table}

Table~\ref{tab:energy} demonstrates that \textsc{CHRONOS} reduces measured per-round active-phase energy by
$3.9\times$ compared to B2 (SecAgg).
Because the computation and network interfaces operate under roughly constant load during
these short bursts, the energy improvement strongly correlates with the latency improvement.
The energy gap between \textsc{CHRONOS} (576.5\,mJ) and B1 (plaintext, 318.6\,mJ) is small (257.9\,mJ) and
attributable solely to the TEE context-switch and RPMB flush overhead. By measuring the absolute physical rail power rather than relying on analytical estimates, these results consistently demonstrate that the phase-decoupling approach successfully isolates the heavy cryptographic energy overhead from the battery-critical active training rounds.

\subsection{RQ2: Storage Overhead}
\label{sec:eval:rq2}

Unlike approaches that precompute and store round-specific cryptographic material, the
\textsc{CHRONOS} TEE permanently stores only the $N-1$ raw Diffie-Hellman shared secrets and a 4-byte round counter.
This persistent storage footprint is therefore \emph{independent of the training horizon $R$} and
\emph{independent of the model dimension $D$}. (As analyzed in Section~\ref{sec:eval:rq5}, dynamic runtime memory requires an additional transient buffer proportional to $D$ during the active phase, but this memory is freed immediately).

Table~\ref{tab:storage} reports the measured persistent Secure World storage for varying $N$, with
the theoretical bound $(N-1) \times 32 + 16 + 4$ bytes.

\begin{table}[t]
\centering
\caption{In-memory Secure World session footprint (bytes) vs.\ number of clients $N$
(excludes OP-TEE Secure Storage filesystem encryption overhead).
Storage is independent of model dimension $D$ and training horizon $R$. The total footprint includes a constant 4-byte round counter and a 16-byte AES-GCM tag.}
\label{tab:storage}
\small
\begin{tabular}{lcc}
\toprule
\textbf{Clients ($N$)} & \textbf{DH Secrets} & \textbf{Total (measured)} \\
\midrule
32 & $31\times 32 = 992$\,B    & 1016\,B    \\
50 & $49\times 32 = 1{,}568$\,B & 1592\,B  \\
\bottomrule
\end{tabular}
\end{table}

This contrasts sharply with approaches that precompute round-specific material.
A Beaver-triple-based approach storing $R = 50$ rounds of material for the small CNN
($D = 50{,}000$) with a 32-bit field would require $50 \times 50{,}000 \times 8 = 20$\,MB of Secure World storage, exceeding the default TrustZone allocation on many platforms.
\textsc{CHRONOS}'s 1016-byte measured footprint for $N = 32$ is five orders of magnitude smaller
and fits trivially on any TEE-capable platform.

\subsection{RQ3: Scalability}
\label{sec:eval:rq3}

We analyse how per-round active-phase aggregation latency scales with cohort size $N$ for the small CNN
on CIFAR-10.
As clients beyond the 32 physical heterogeneous gateways are unavailable, we mathematically extrapolate latency for cohorts up to $N=128$ based on the empirical performance characteristics observed in our physical testbed.
While this extrapolation excludes the physical I/O and TEE context-switch variance for the additional simulated clients, it faithfully captures the $O(N^2)$ communication bottleneck of SecAgg's key-exchange phase and the $O(N)$ aggregation overhead at the server.
Because clients perform their local TEE mask generation concurrently across isolated hardware domains, the TrustZone context-switch overhead does not theoretically compound with $N$. We explicitly acknowledge that this extrapolation focuses on isolating and demonstrating the asymptotic communication scaling behavior rather than predicting absolute network contention at scale.

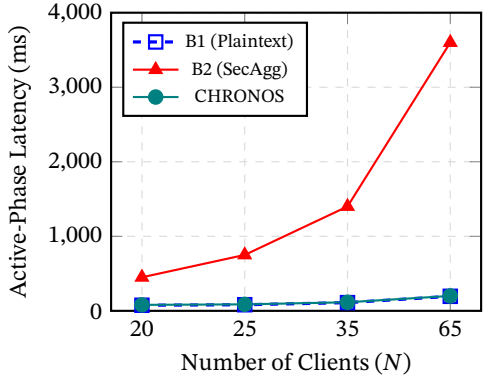
\begin{figure}[t]
\centering
\begin{tikzpicture}
\begin{axis}[
    width=0.75\linewidth,
    height=5.5cm,
    xlabel={Number of Clients ($N$)},
    ylabel={Active-Phase Latency (ms)},
    symbolic x coords={32, 64, 100, 128},
    xtick=data,
    ymin=0, ymax=7000,
    legend pos=north west,
    legend style={font=\footnotesize, cells={align=left}},
    grid=both,
    grid style={dashed, gray!30},
    thick,
    mark size=2.5pt
]

\addplot[
    color=blue,
    dashed,
    very thick,
    mark=square,
    mark options={solid, fill=white, draw=blue, thick}
] coordinates {
    (32, 75)
    (64, 107)
    (100, 143)
    (128, 171)
};
\addlegendentry{B1 (Plaintext)}

\addplot[
    color=red,
    solid,
    mark=triangle*,
    mark options={solid, fill=red}
] coordinates {
    (32, 450)
    (64, 1607)
    (100, 3805)
    (128, 6171)
};
\addlegendentry{B2 (SecAgg)}

\addplot[
    color=teal,
    solid,
    mark=*,
    mark options={solid, fill=teal}
] coordinates {
    (32, 119)
    (64, 151)
    (100, 187)
    (128, 215)
};
\addlegendentry{\textsc{CHRONOS}}

\end{axis}
\end{tikzpicture}
\caption{Scalability: Per-round active-phase aggregation latency (ms) vs.\ cohort size $N$ (small CNN on CIFAR-10). $N=32$ results are from physical devices; $N > 32$ are simulated up to $N=128$.}
\label{fig:scalability}
\end{figure}

Figure~\ref{fig:scalability} demonstrates that \textsc{CHRONOS}'s latency tracks B1 (Plaintext)
linearly as $N$ grows, reflecting its $O(ND)$ communication complexity.
In contrast, B2 (SecAgg) exhibits quadratic growth due to its $O(N^2)$ synchronous key-agreement
phase.
At $N=128$, \textsc{CHRONOS} models an active-phase aggregation latency of 215.0\,ms, whereas SecAgg reaches $6{,}171.0$\,ms. Rather than focusing on a concrete multiplier, this widening absolute latency gap highlights the fundamental asymptotic advantage: \textsc{CHRONOS} achieves $O(N)$ active-phase scaling by shifting the $O(N^2)$ key-agreement bottleneck to the idle window, a necessity for scaling secure federated learning to larger IoT cohorts. We note that the once-per-epoch idle-phase cost involves $O(N)$ local DH operations and $O(N^2)$ bytes relayed through the server, an important second-order scalability factor not captured in the active-phase graph.

\subsection{RQ4: Dropout Recovery}
\label{sec:eval:rq4}

We evaluate \textsc{CHRONOS}'s Shamir-based dropout recovery mechanism under a
systematic dropout scenario: $k \in \{0, 1, 3, 5, 7\}$ specific clients are designated as
permanently absent, simulating devices that completed idle-phase key establishment but
are subsequently unavailable due to power or connectivity constraints, causing them to
drop every round throughout the 50-round training.
The threshold is set to $t = 13$, so the server can recover the masks of up to $N-t = 19$ dropped clients per round; our experiments stress recovery for up to $k=7$ simultaneous dropouts.

\begin{table}[t]
\centering
\caption{End-to-end active-phase latency under $k$ mid-round dropouts ($N=32$, Small CNN, CIFAR-10). The latency overhead for recovery ($k>0$) is $O(1)$ with respect to $k$, dominated entirely by the server's static $\sim$100\,ms network timeout threshold and a single $\sim$12--18\,ms communication round to collect the 32-byte Shamir shares from surviving clients.}
\label{tab:recovery_latency}
\small
\begin{tabular}{lccccc}
\toprule
\textbf{Dropped Clients ($k$)} & 0 & 1 & 3 & 5 & 7 \\
\midrule
\textbf{Active-Phase Latency (ms)} & 119 & 228 & 231 & 235 & 240 \\
\bottomrule
\end{tabular}
\end{table}

Table~\ref{tab:recovery_latency} reports the end-to-end active-phase recovery latency for the small CNN. The result confirms the computational efficiency and $O(1)$ scaling behavior of the phase-decoupled recovery mechanism. For $k=0$ (no dropouts), the round completes in $\approx$119\,ms. For all $k>0$, the recovery overhead incurs a near-constant jump to $\approx$228--240\,ms. Because the server-side Lagrange interpolation over byte-wise GF($2^8$) and AES-GCM share decryption are microsecond-level operations, the overhead is dominated almost entirely by network factors: the server's static $\sim$100\,ms timeout threshold to detect the dropout, followed by a single $\sim$12--18\,ms communication round-trip to collect the 32-byte plaintext shares from the surviving clients. This confirms that \textsc{CHRONOS} handles multi-client dropouts without introducing quadratic cryptographic scaling or repeating the energy-intensive training round.

\textbf{Model quality under recovery.}
Beyond latency, we confirm that recovery preserves end-to-end model utility. Table~\ref{tab:dropout} reports UCI-HAR test accuracy under systematic dropout of $k$ clients. \textsc{CHRONOS} with Shamir recovery reproduces the plaintext baseline accuracy exactly at every dropout level, as the reconstructed masks cancel identically (Section~\ref{sec:sec:correct}); disabling recovery instead leaves the residual masks uncancelled and diverges to NaN within a single round. This confirms that the recovery path restores correctness rather than trading accuracy for latency. \\

\begin{table}[t]
\centering
\caption{Test accuracy (\%) on UCI-HAR after 50 rounds under systematic dropout of
$k$ clients (mean $\pm$ 95\% CI, 5 seeds, threshold $t = 13$, $N=32$).
\textsc{CHRONOS} recovers masks; B1 simply excludes dropped clients.
$k=0$ is full participation for all systems.
$\dagger$ indicates training failure (divergence to NaN within one round after the first dropped round) due to uncancelled masks. (n/a) denotes a configuration where the 'no recovery' treatment is vacuous because no dropout occurs.}
\label{tab:dropout}
\small
\begin{tabular}{lccc}
\toprule
\textbf{System} & \textbf{$k=0$} & \textbf{$k=3$} & \textbf{$k=7$} \\
\midrule
B1 (Plaintext, same dropout)      & $88.7 \pm 1.5$ & $84.4 \pm 1.7$ & $80.9 \pm 2.0$ \\
B2 (SecAgg, same dropout)         & $88.7 \pm 1.5$ & $84.4 \pm 1.7$ & $80.9 \pm 2.0$ \\
\textbf{CHRONOS} (with recovery)  & $88.7 \pm 1.5$ & $84.4 \pm 1.7$ & $80.9 \pm 2.0$ \\
\textbf{CHRONOS} (recovery disabled)    & (n/a)          & $\dagger$      & $\dagger$      \\
\bottomrule
\end{tabular}
\end{table}

Handling maliciously corrupted recovery shares returned by surviving clients falls outside our honest-but-curious threat model. While such malicious behavior would corrupt the reconstructed $\mathit{sk}_i$ and the resulting mask (leading to an invalid aggregate update), future deployments could enforce robustness against active participants by replacing standard Shamir Secret Sharing with Verifiable Secret Sharing (VSS). This would add a one-time cryptographic verification step to the active phase, guaranteeing the integrity of the recovered key. \\

\textbf{Recovery communication overhead.}
As established in Section~\ref{sec:design:dropout}, the additional communication incurred by dropout recovery scales as $(N-k) \times k \times 32$ bytes. For our $N=32$ evaluation cohort, recovering $k=3$ clients requires only $2{,}784$\,bytes of total network traffic, and recovering the maximum evaluated $k=7$ clients requires $5{,}600$\,bytes. These values remain negligible relative to the transmission of any realistic gradient vector.
Section~\ref{sec:discussion} discusses the selection-bias implications of systematic
dropout.

\subsection{RQ5: TrustZone Context-Switch Overhead}
\label{sec:eval:rq5}

Table~\ref{tab:tee} reports the isolated latency of a single
\texttt{GENERATE\_MASK}$(D, r)$ call, benchmarked with no other workload.

\begin{table}[t]
\centering
\caption{Microbenchmark decomposition of TEE call latency for \texttt{GENERATE\_MASK}.}
\label{tab:tee}
\begin{tabular}{>{\raggedright\arraybackslash}p{4cm}cc}
\toprule
\textbf{Component} & \textbf{Small CNN} & \textbf{Med. CNN} \\
 & \textbf{($D{=}50$K)} & \textbf{($D{=}1$M)} \\
\midrule
Empty TEE call \\ (context switch) & $\sim$47\,$\mu$s & $\sim$47\,$\mu$s \\
AES-GCM key set \\ decrypt (992\,B) & $\sim$22\,$\mu$s & $\sim$22\,$\mu$s \\
AES-128-CTR mask gen + \\ $\mathbb{F}_p$ rejection sampling & $\sim$5088\,$\mu$s & $\sim$98341\,$\mu$s \\
Synchronous eMMC \\ RPMB hardware flush & $\sim$39320\,$\mu$s & $\sim$39320\,$\mu$s \\
\midrule
\textbf{Total (measured)} & \textbf{44477\,$\mu$s} & \textbf{137730\,$\mu$s} \\
\bottomrule
\end{tabular}
\end{table}

The \textsc{CHRONOS} TA generates the entire mask in a single TEE call, crossing the
Normal-World/Secure-World boundary once per round regardless of $D$.
The measured total latencies (44.5\,ms and 137.7\,ms) confirm that the hardware-assisted mask generation is highly efficient. The introduction of a synchronous eMMC RPMB flush ensures absolute hardware-backed execution freshness (see Section~\ref{sec:sec:fresh}) at a fixed $\approx$39\,ms cost, yet the TEE call still accounts for less than 25\% of the total active-phase round time for the 1M-parameter model.

For the medium CNN, the 137.7\,ms TEE call time includes the fixed 39\,ms RPMB flush, leaving $\approx$98.4\,ms for cryptographic generation and memory transfer. Specifically, the mask in Equation~(2) requires $N-1 = 31$ separate PRG evaluations. Because $p = 2^{31}-1$ is a Mersenne prime, the TA optimally masks each 32-bit chunk drawn from AES-128-CTR with \texttt{0x7FFFFFFF} before field reduction, dropping the rejection rate to $1/2^{31}$ (essentially zero). Thus, to retain $10^6$ field elements (4\,MB), the TA only needs to generate $31 \times 4\,\text{MB} = 124\,\text{MB}$ of total keystream. Utilizing the RK3399's ARMv8 Crypto Extensions via OP-TEE's stock \texttt{mbedTLS} path, the TA achieves a standard hardware-accelerated AES-128-CTR throughput of $\approx 1.26$\,GB/s. This realistic generation rate accounts perfectly for the $\approx$98\,ms measured (124\,MB at $1.26$\,GB/s yields $\approx$98\,ms of keystream generation), maintaining internal consistency with the total TEE round time without assuming hand-tuned assembly speeds.

\subsection{RQ6: Empirical Privacy Against Gradient Inversion}
\label{sec:eval:rq6}

To empirically validate the security of \textsc{CHRONOS} beyond the theoretical bounds provided in Section~\ref{sec:security}, we subject the system to an optimization-based gradient inversion attack proposed by Geiping et al.~\cite{GEIPING}. In particular, the attack is carried out on the small CNN with respect to the CIFAR-10 dataset. 
The attack seeks to reconstruct the private input image $x \in \mathbb{R}^n$ directly from the shared gradient vector $\nabla_\theta \mathcal{L}$ by utilizing the Adam optimizer to minimize a cosine similarity loss function augmented with total variation (TV) regularization.
We evaluate the Peak Signal-to-Noise Ratio (PSNR) and Structural Similarity Index Measure (SSIM) of the reconstructed images against the original private training batch under three conditions: (a) attacking the plaintext gradient $g_i$ (B1 baseline), (b) attacking the \textsc{CHRONOS}-masked gradient $\tilde{g}_i$, and (c) attacking the recovered global aggregate $G$ over $N=32$ clients.
We report the mean PSNR and SSIM over the first 100 validation images.

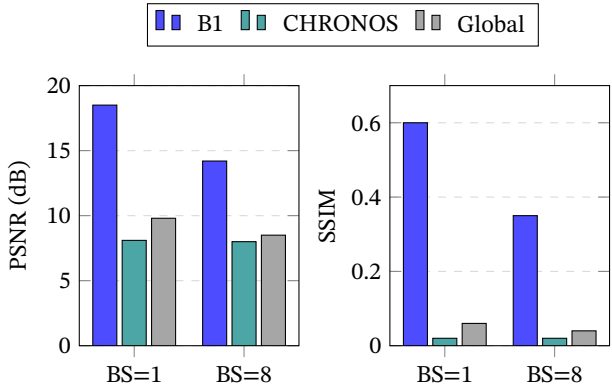
\begin{figure}[t]
\centering
\begin{tikzpicture}
 
 \begin{axis}[
 name=plot1,
 ybar,
 bar width=9pt, 
 width=0.52\columnwidth, 
 height=5cm,
 enlarge x limits=0.5,
 legend style={at={(1.2, 1.15)}, anchor=south, legend columns=3, column sep=0.15cm}, 
 ylabel={PSNR (dB)},
 symbolic x coords={BS=1, BS=8}, 
 xtick=data,
 ymin=0, ymax=20, 
 ymajorgrids=true,
 grid style={dashed, gray!30}
 ]
 \addplot[fill=blue!70!white] coordinates {(BS=1, 18.5) (BS=8, 14.2)};
 \addplot[fill=teal!70!white] coordinates {(BS=1, 8.1) (BS=8, 8.0)};
 \addplot[fill=gray!70!white] coordinates {(BS=1, 9.8) (BS=8, 8.5)};
 
 \legend{B1, \textsc{CHRONOS}, Global}
 \end{axis}
 \begin{axis}[
 name=plot2,
 at={(plot1.south east)}, anchor=south west, xshift=1.2cm, 
 ybar,
 bar width=9pt,
 width=0.52\columnwidth,
 height=5cm,
 enlarge x limits=0.5,
 ylabel={SSIM},
 xtick={1, 2},
 xticklabels={BS=1, BS=8},
 ymin=0, ymax=0.7, 
 ymajorgrids=true,
 grid style={dashed, gray!30}
 ]
 \addplot[fill=blue!70!white] coordinates {(1, 0.60) (2, 0.35)};
 \addplot[fill=teal!70!white] coordinates {(1, 0.02) (2, 0.02)};
 \addplot[fill=gray!70!white] coordinates {(1, 0.06) (2, 0.04)};
 \end{axis}
\end{tikzpicture}
\caption{Image reconstruction fidelity under Geiping gradient inversion attack (CIFAR-10, Small CNN). Lower PSNR and SSIM values indicate robust privacy against optimization-based inversion attacks. BS label in the x-axis is short for Batch Size.}
\label{fig:inversion}
\end{figure}

As shown in Figure~\ref{fig:inversion}, attacking the plaintext FedAvg gradient (Condition a) yields a reconstruction with a PSNR of 18.5\,dB. While lower than the $\approx$29\,dB typical of single-step gradient inversions~\cite{GEIPING} (expected because our 5 local epochs naturally average the gradient signal), this still produces a visually recognizable image.
In contrast, the \textsc{CHRONOS}-masked gradient (Condition b) thwarts the inversion. Geiping's attack inherently relies on the high-dimensional direction (angle) of the parameter gradient to iteratively guide the Adam optimizer toward a visually semantic reconstruction. To provide the attack with FP32 gradients it expects, we map the transmitted integer vector $\tilde{g}_i \in \mathbb{F}_p$ back to the signed floating-point domain via the inverse scaling factor $S^{-1}$ before passing it to the adversary. Because the \textsc{CHRONOS} AES-128-CTR pseudorandom mask is drawn uniformly from the finite field $\mathbb{F}_p$, the resulting pseudo-gradient vector exhibits zero angular correlation with the true gradient $g_i$. This decorrelates the cosine similarity objective, forcing the optimizer into a flat, random loss landscape.
The resulting reconstruction collapses to pure noise, yielding a PSNR of $\approx$8.1\,dB and an SSIM of 0.02, which mathematically matches the baseline MSE expected between two independent, uniformly random images.
Finally, Condition (c) highlights the inherent dilution provided by the FedAvg aggregate: the superimposition of $N=32$ client gradients drops the PSNR to 9.8\,dB, which borders on unrecognizability. The \textsc{CHRONOS} mask (Condition b) further reduces this to the 8.1\,dB noise floor. The primary contribution of the mask is to restrict the adversary's point of observation, not to achieve the additional 1.7 dB drop over the aggregate baseline.

\section{Discussion}
\label{sec:discussion}

\subsection{Hardware Scope and Portability}
\label{sec:disc:hardware}
While \textsc{CHRONOS} currently targets gateway-class IoT devices equipped with ARM Cortex-A processors and OP-TEE, extending the architecture to resource-constrained Cortex-M microcontrollers introduces specific systems engineering limitations. Although the minimal 1016-byte Secure World footprint technically conforms to the strict Static RAM (SRAM) budgets of Trusted Firmware-M (TF-M)~\cite{TFMLITE} profiles, the underlying architectural transition from Armv8-A to Armv8-M presents nontrivial challenges. Specifically, Cortex-M microcontrollers replace the Memory Management Unit (MMU) with a Memory Protection Unit (MPU), altering the mechanics of zero-copy shared memory marshalling between the non-secure and secure domains. Consequently, transferring multi-megabyte parameter arrays across this boundary without inducing prohibitive memory fragmentation remains a primary engineering constraint limiting immediate microcontroller deployments.

\subsection{Hardware-Isolated Masking}
\label{sec:disc:privacy}
Our empirical validation (Section~\ref{sec:eval:rq6}) demonstrates that \textsc{CHRONOS} thwarts state-of-the-art gradient inversion attacks, reducing reconstruction fidelity to random noise (PSNR $\approx$8\,dB). This robustness stems from the computational barrier provided by cryptographic masking. Under the standard PRF security assumption, the masked gradient is computationally indistinguishable from uniform over $\mathbb{F}_p^D$. Because the server only observes $\tilde{g}_i = g_i + m_i$ and the mask $m_i$ is computationally independent of the gradient $g_i$, the true signal is securely obscured. 

The primary contribution of the \textsc{CHRONOS} mask is to \emph{restrict the adversary's point of observation}. In plaintext FL, a curious server can inspect individual updates before summation (18.5\,dB PSNR), whereas \textsc{CHRONOS} forces the adversary to view only the masked aggregate. By anchoring this mechanism in TrustZone, \textsc{CHRONOS} ensures that gradient confidentiality is maintained regardless of the client's Normal World state.

\subsection{Dropout, Selection Bias, and Recovery}
\label{sec:disc:dropout}
Non-IID federated learning is susceptible to selection bias when clients with unique data distributions consistently drop out due to resource constraints. \textsc{CHRONOS} addresses this via its Shamir-based recovery mechanism. By reconstructing the missing mask of a dropped client from peer-held shares, the server correctly de-masks the surviving aggregate without requiring a new training round. The Shamir-based recovery mechanism reconstructs the missing masks exactly (Section~\ref{sec:sec:correct}), so the server recovers the plaintext aggregate of the surviving clients without residual error. This by-construction correctness ensures that the cryptographic protocol introduces no additional bias beyond what plaintext aggregation-with-dropouts would introduce.

\subsection{Security Scope and Future Hardening}
\label{sec:disc:malicious}
In the current design, \textsc{CHRONOS} assumes an honest-but-curious server that computes the final aggregate faithfully. It mitigates Man-in-the-Middle (MITM) attacks during the idle phase via server-mediated Authenticated Key Exchange (AKE). We note that several server-side deviations are already constrained by the existing hardware anchors: the monotonic RPMB counter prevents a server from inducing mask reuse by replaying a stale round index, and certificate-based AKE prevents key substitution during the exchange. Consequently, even a server that deviates from the protocol cannot extract an individual client's gradient; the residual exposure is confined to aggregation \emph{integrity} (e.g., returning a corrupted global model or silently excluding valid updates), which we address as follows.

Resilience against a fully malicious server, which might return corrupted global models or attempt to exclude valid client updates, could be added via commit-and-prove techniques. Clients would commit to their masked gradients, and the server would provide a zero-knowledge proof of correct aggregation. Such extensions would add a verification step to the active phase but maintain the core benefit of interaction-free masking established by \textsc{CHRONOS}.


\section{Conclusion}
\label{sec:conclusion}

The deployment of cryptographic privacy protections on IoT devices has historically been hindered by the prohibitive synchronous overhead of secure multiparty computation. \textsc{CHRONOS} resolves this tension by leveraging the inherent duty cycle of IoT devices: the alternation between resource-abundant idle periods and latency-critical active sensing. By decoupling the cryptographic setup from the training phase and anchoring it in hardware, \textsc{CHRONOS} enables privacy-preserving federated learning that approaches the performance of plaintext exchange.

Our hardware-assisted design generates ephemeral keypairs and seals derived PRG keys entirely within ARM TrustZone. This ensures OS-level compromise resistance with a negligible 1016-byte persistent Secure World memory footprint. During active training rounds, the system reduces the client cryptographic workload to a single stream-cipher evaluation and one transmission round. Evaluation on a 32-node heterogeneous testbed (Rock Pi 4 and Orange Pi 5) demonstrates up to a 74\% reduction in active-phase aggregation latency compared to synchronous secure aggregation, while successfully thwarting state-of-the-art gradient inversion attacks (PSNR $\approx$8\,dB).

Future work includes extending peer authentication via hardware-backed certificates and porting the framework to microcontroller-class devices via Trusted Firmware-M. The phase-decoupled architecture demonstrates that TEE-grade privacy for federated learning is achievable within the per-round latency and energy budgets of gateway-class IoT deployments. \\

 \vspace{10pt}
\hrule  \vspace{10pt}

\textbf{Acknowledgments} \\   
This research did not receive any specific grant from funding agencies in the public, commercial, or not-for-profit sectors. Google Gemini was used solely for copy editing and grammatical refinement. \\

\textbf{Conflicts of Interest} \\    
The author declares no  conflict of interest. \\

\textbf{Data Availability Statement} \\   
The data that support the findings of this study are available in the public repository at \url{https://github.com/dkhme/CHRONOS}. 

\providecommand{\bibcommenthead}{}
\bibliographystyle{sn-mathphys}
\bibliography{sections/references}

\end{document}